\documentclass[  letterpaper, superscriptaddress, nofootinbib, onecolumn, prepint]{revtex4}
\pdfoutput=1
\usepackage{graphicx}
\usepackage{microtype}
\usepackage{amsmath}
\usepackage{amssymb}
\usepackage{subfigure}
\usepackage{hyperref}
\usepackage{url}
\usepackage{xcolor}
\usepackage{color}
\usepackage{mathrsfs}
\usepackage{lmodern}
\usepackage{calrsfs}
\usepackage{amsfonts}
\usepackage{eufrak}
\usepackage{array}
\usepackage{tabularx}
\usepackage{eucal}
\usepackage{latexsym}
\usepackage{ragged2e}
\usepackage{epsfig}
\usepackage{textcomp}
\usepackage{setspace}

\usepackage{bm, braket}

\usepackage{caption}
\DeclareCaptionJustification{justified}{\leftskip=0pt \rightskip=0pt \parfillskip=0pt plus 1fil}
\definecolor{vividviolet}{rgb}{0.62, 0.0, 1.0}
\definecolor{amaranth}{rgb}{0.9, 0.17, 0.31}
\definecolor{palatinateblue}{rgb}{0.15, 0.23, 0.89}
\definecolor{brightpink}{rgb}{1.0, 0.0, 0.5}
\definecolor{cornflowerblue}{rgb}{0.39, 0.58, 0.93}
\definecolor{deepcarminepink}{rgb}{0.94, 0.19, 0.22}
\definecolor{radicalred}{rgb}{1.0, 0.21, 0.37}

\hypersetup{ linktoc=all,
	colorlinks, linkcolor={palatinateblue},
	citecolor={brightpink}, urlcolor={amaranth}
}

\renewcommand{\d}[1]{\ensuremath{\operatorname{d}\!{#1}}}

\graphicspath{{Images/}}
\renewcommand{\d}[1]{\ensuremath{\operatorname{d}\!{#1}}}
\makeatletter
\def\@fnsymbol#1{\ensuremath{\ifcase#1\or \ddagger \or  $\textleaf$  \or \dagger
		\else\@ctrerr\fi}}%

\makeatother

\def\sideremark#1{\ifvmode\leavevmode\fi\vadjust{\vbox to0pt{\vss
			\hbox to 0pt{\hskip\hsize\hskip1em
				\vbox{\hsize1.3cm\tiny\raggedright\pretolerance10000
					\noindent #1\hfill}\hss}\vbox to8pt{\vfil}\vss}}}%
\def\beq{\begin{equation}}
\def\eeq{\end{equation}}

\begin{document}
	
	\title{On the Curvature Invariants of the Massive Banados-Teitelboim-Zanelli Black Holes and Their Holographic Pictures}

	\author{Mahdis \surname{Ghodrati}}
	\email{mahdisg@yzu.edu.cn}
	\affiliation{Center for Gravitation and Cosmology, College of Physical Science and Technology, Yangzhou University, \\180 Siwangting Road, Yangzhou City, Jiangsu Province, P.R. China 225002}
	\affiliation{School of Aeronautics and Astronautics, Shanghai Jiao Tong University, Shanghai 200240, China}
	
	\author{Daniele \surname{Gregoris}}
	\email{danielegregoris@libero.it}
	\affiliation{Center for Gravitation and Cosmology, College of Physical Science and Technology, Yangzhou University, \\180 Siwangting Road, Yangzhou City, Jiangsu Province, P.R. China 225002}
	\affiliation{School of Aeronautics and Astronautics, Shanghai Jiao Tong University, Shanghai 200240, China}

\begin{abstract}
{In this paper, the curvature structure of a (2+1)-dimensional black hole in the massive-charged-Born-Infeld gravity is investigated. The metric that we consider is characterized by four degrees of freedom which are the mass and electric charge of the black hole, the mass of the graviton field, and a cosmological constant. For the charged and neutral cases separately, we present various constraints among scalar polynomial curvature invariants which could invariantly characterize our desired spacetimes. Specially, an appropriate  scalar polynomial curvature invariant and a Cartan curvature invariant which together could {\it detect} the black hole horizon would be explicitly constructed. Using algorithms related to the focusing properties of a bundle of light rays on the horizon which are accounted for by the Raychaudhuri equation, a procedure for isolating the black hole parameters, as the algebraic combinations involving the curvature invariants, would be presented. It will be shown that this technique could specially be applied for black holes with zero electric charge, contrary to the cases of solutions of lower-dimensional non-massive gravity. In addition, for the case of massive (2+1)-dimensional black hole, the irreducible mass, which quantifies the maximum amount of energy which could be extracted from a black hole would be derived. Therefore, we show that the Hawking temperatures of these black holes could be reduced to the pure curvature properties of the spacetimes. Finally, we comment on the relationship between our analysis and the novel roles it could play in numerical quark-gluon plasma simulations and other QCD models and also black hole information paradox where the holographic correspondence could be exploited.}
\end{abstract}

\maketitle

\section{Introduction}

The analysis of quark-gluon plasmas play a cornerstone role in studying many physical phenomena ranging from studies of neutron stars\footnote{The typical mass, radius, and temperature of a neutron star are respectively $10^{33}$ g, 10 km, and $10^{12}$ K.}, early universe,  earth-based experiments  such as RHIC  at Brookhaven National Laboratory and LHC at CERN, and many others.

Due to the strongly coupled nature of  quantum-chromodynamics theories, theoretical predictions rely on numerical lattice simulations, diagrammatic expansions, and effective field theory. 
Fortunately, holography, the gauge/gravity duality and  specifically the AdS/QCD correspondence could allow us to study strongly coupled systems arising in $d$-dimensional boundary gauge theories as gravitational systems in $d+1$-dimensional bulk spacetimes.

In the AdS/CFT setup, for studying thermalization problems in the CFT side, one needs to consider gravity background solutions with a black hole. 

For this purpose, various black hole solutions in different dimensions could be used.  Actually, beginning from the Banados-Teitelboim-Zanelli (BTZ) black hole being introduced in $(2+1)$ dimensions, a number of mathematical solutions of the Einstein equations in three-dimensional gravity has been derived.

The case of three-dimensional gravity is in fact a special one. The core difference with the case of $(3+1)$ and higher-dimensional gravity theories is that the Weyl part of the curvature is trivial, and therefore the lower-dimensional black hole solutions could only exist for spacetimes supported by a negative cosmological constant. For the case of three-dimensional gravity, in order to have non-trivial dynamics, one needs to add additional degrees of freedom. Specifically, the (2+1)-dimensional Einstein gravity could be augmented with the introduction of a new degree of freedom which could be interpreted as a "massive spin-2 graviton" into a new family of {\it three-dimensional massive gauge theories} \cite{Deser:1982vy,Deser:1981wh,Deser:1983tn}.

Then, the analysis of the linearized regimes of this solution would reveal that the sign in front of the Ricci scalar must be flipped in order to get a solution with massive graviton and with positive energy \cite{Carlip:2008jk}.

The solution describing an electrically charged black hole in the coupled \textit{ "Maxwell-Topologically-Massive gravity"} has first been proposed in \cite{Hendi:2016pvx}.  Note that in fact, these massive gravity theories do not have any Boulware-Deser ghost \cite{Hassan:2011hr} and as Vegh showed in \cite{Vegh:2013sk}, the graviton in such theories behave like a lattice and it could then exhibit a Drude peak. Also, as for the charged case it is the generalization of Maxwell theory to nonlinear electrodynamics, there would not be any singularity of electromagnetic field, or singularities of D-branes present in these theories \cite{Cai:2014znn, Leigh:1989jq,Wiltshire:1988uq}. Therefore, this theory and its various solutions could in fact be used to study various QCD and condensed matter systems.

Additionally, many holographic features of these models, such as charged case, existence of Van der Waals like behaviors, nonlinear electrodynamic features, thermodynamic properties, metal/superconductor phase transitions, etc, have been studied in various works, see \cite{Cai:2014znn, Vegh:2013sk, Hu:2015dnl,Zeng:2015tfj,Ghodrati:2019hnn,Zhou:2019jlh,Ghodrati:2016ggy} as a few examples.

In this paper, by applying the concept of {\it irreducible mass} which has been proposed by Christodoulou and Ruffini \cite{Christodoulou:1970wf,Christodoulou:1972kt}, we investigate how the mass of the graviton field would affect the maximum amount of energy which could be extracted from a (2+1)-dimensional black hole, and then using holography we evaluate how the dual momentum dissipations would affect QCD phases. In addition, these results could further examine various proposals for the black holes information paradox.

Now the main problem is that in all the applications of numerical techniques for simulating quark-gluon plasma with high-resolution shock-capturing methods which is being used as the prototype, the procedure would be affected by the method of determining the causally-disconnected spacetime regions which should be removed from the quantitative results \cite{Rezzolla:1981wh}. These regions are those which live outside the light-cone.  The same problem would also arise in the simulations of binary black holes in numerical relativity, which are being used for extracting information about the gravitational waves spectra, such as recent measurements of LIGO \cite{Abbott:2016nmj}.

So the most important task in dealing with these black holes is determining exactly their event horizon, which would be a codimension one null hypersurface, where from there the causal geodesics would not reach to the future null infinity. Since the event horizon is very non-local, one needs the knowledge of the full spacetime in order to locate it.

A number of excision techniques in numerical relativity have been developed for removing the spacetime region inside black hole horizons \cite{Baumgarte:2010ndz}. These techniques are actually computationally expensive, however a novel method which has been first speculated by Abdelqader and Lake, in \cite{Abdelqader:2014vaa} and then rigorously proved in \cite{Page:2015aia, McNutt:2017gjg}, showed that the location of a stationary black hole horizon could be found from the algebraic equations involving only some scalar curvature invariants \cite{Page:2015aia,McNutt:2017paq}. Even some other properties of the black hole such as its mass and spin could also be determined by this method.

In particular,  two methods have been explored in this line, one involving polynomial scalar curvature invariants, and a second one which deals with the Cartan curvature invariants.  It has been shown that the second method is more efficient computationally, as it relies on the first degree foliation-independent quantities. Also, it has been shown that both of these methods would fail for the lower-dimensional general relativity black hole solutions unless a non-trivial matter-energy tensor would be added \cite{Gregoris:2019ycf} to the action.

In this paper, one of our main results is to show that, for the case of the solutions of  massive gravity where the graviton has a mass parameter, the applicability of the method involving polynomial scalar curvature invariants could be recovered, even for black holes with no electric charge.

So this paper is organized as follows. In section (\ref{sec:s1}), we will construct a polynomial scalar curvature invariant and a Cartan curvature invariant which together could {\it locate} the horizon of the (2+1)-dimensional black hole in our specific chosen massive gravity theory which has a lot of applications in AdS/CMT \cite{Hendi:2016pvx,Hendi:2017vyz}, QCD \cite{Hendi:2017ibm}, quantum information \cite{Ghodrati:2019hnn} and studies of black hole information paradox.

Then, we compare the results derived in that section with the two other mentioned methods and also we compare those results with the ones from other methods in the literature. Moreover, in section two, we will also exhibit a number of relationships between the curvature components of this black hole solution, which enlighten the different roles played by the electric charge and also by the mass of the graviton field, in shaping the specific properties of the black hole solutions in these gravity backgrounds. 

In section (\ref{s3}), we will show a procedure which demonstrated to be a constructive method which could also preserve locality and it would be used for inferring the values of the parameters of the black hole using appropriate algebraic combinations of the curvature quantities.

Then, in section (\ref{s4}) we apply the Cartan method for locating the black hole horizon starting from the Raychaudhuri equation which accounts for the focusing properties of a bundle of light rays on the horizon and then we  compare the results with those derived from the previous method. 

Next, in section (\ref{s5}) we compute the irreducible mass of the black hole and comment on its physical significance and how the graviton mass parameter would affect it.

In section (\ref{sec:holographyDisc}), we discuss the dual holographic pictures of the curvature invariants in the bulk, specifically for the case of massive gravity theory, from the boundary CFT perspective.

 Finally, in section (\ref{s6}) we comment on our results and conclude with discussions of future directions.

\section{Locating the event horizon in terms of the zeroes of curvature invariants}\label{sec:s1}

In this section we first investigate the problem of locating the horizon of the Banados-Teitelboim-Zanelli (BTZ) black hole solution of massive gravity using the appropriate curvature quantities.

In particular, we focus on two specific methods which deal with the scalar polynomial and Cartan curvature invariants respectively. These two algorithms have already been proved to constitute valuable techniques for locating the horizons of astrophysical black holes, higher-dimensional black holes of string theory and also lower dimensional black holes in (2+1) and (1+1)-gravity theories, \cite{Page:2015aia,McNutt:2017paq,Gregoris:2019ycf,McNutt:2017gjg,Coley:2017woz,Coley:2017vxb,McNutt:2018fjn,Coley:2019ylo}.

Here, the word {\it locate} means that one could derive certain curvature quantities which would vanish on the black hole horizon and only there and therefore provide sharp and precise information on its location without delivering any false positive.

While, the existence of these quantities could generally be proved mathematically, as long as the spacetime admits a stationary horizon, their specific constructions must be performed explicitly, case by case, for every solution. As a side result, we also present a number of constraints between the components of the curvature invariants which show the geometrical differences between various black hole models.

The theory that we consider would be the three dimensional Einstein-massive gravity with the following action \cite{Hendi:2016pvx}
\begin{equation}
\mathcal{L}=-\frac{1}{16\pi }\int d^{3}x\sqrt{-g}\left[R-2\Lambda+L(\mathcal{F})+M^{2}\sum_{i}^{4}c_{i}\mathcal{U}_{i}(g,h)\right],
\label{Action}
\end{equation}%
where $R$ is the scalar curvature, $L(\mathcal{F})$ is an arbitrary Lagrangian of electrodynamics and $\Lambda$ is the cosmological constant.

The fixed symmetric tensor satisfies the relation $h_{\mu \nu }=diag(0,0,c_0^{2}h_{ij})$ and the corresponding symmetric polynomials  $\mathcal{U}_{i}$ could be written as\footnote{In this paper we change the radial coordinate as $r \to \frac{1}{z}$.} $\mathcal{U}_{1}=c_0/r$ and also we set $\mathcal{U}_{2}=\mathcal{U}_{3}=\mathcal{U}_{4}=0$

The symmetric polynomials of the eigenvalues of the $d\times d$ matrix $\mathcal{K}_{\nu }^{\mu }=\sqrt{%
g^{\mu \alpha }h_{\alpha \nu }}$, for any symmetric tensor could be written as
\begin{eqnarray}\label{eq-Ui}
\mathcal{U}_{1} &=&\left[ \mathcal{K}\right] ,\;\;\;\;\;\mathcal{U}_{2}=%
\left[ \mathcal{K}\right] ^{2}-\left[ \mathcal{K}^{2}\right] ,\;\;\;\;\;%
\mathcal{U}_{3}=\left[ \mathcal{K}\right] ^{3}-3\left[ \mathcal{K}\right] %
\left[ \mathcal{K}^{2}\right] +2\left[ \mathcal{K}^{3}\right] ,  \notag \\
&&\mathcal{U}_{4}=\left[ \mathcal{K}\right] ^{4}-6\left[ \mathcal{K}^{2}%
\right] \left[ \mathcal{K}\right] ^{2}+8\left[ \mathcal{K}^{3}\right] \left[
\mathcal{K}\right] +3\left[ \mathcal{K}^{2}\right] ^{2}-6\left[ \mathcal{K}%
^{4}\right].
\end{eqnarray}

The field equations of the massive gravity theory, in which the black hole solution we will study arises, can be written symbolically as
\beq
G_{ab}+\Lambda g_{ab}+M^2 \chi_{ab}=8 \pi T_{ab}\,,
\eeq
where on the right hand side there is the stress-energy tensor, a function of $L({\mathcal F})$ and of its first derivative,  accounting for the matter content. The crucial difference with respect to Einstein gravity appears instead in the left hand side due to the presence of a term sensitive to the mass of the graviton field  in which the tensor $\chi_{ab}$ is a certain function of the ${\mathcal U}_i$ derived from the Lagrangian (\ref{Action}) by applying a variational approach. We refer the reader to \cite{Hendi:2016pvx} for the explicit expressions of the tensors $\chi_{ab}$ and $T_{ab}$.

Therefore, in this theory we can apply the cosmologist way of thinking which re-absorbs the cosmological constant into the stress-energy tensor re-interpreting it as a sort of dark energy. Thus, the mass of the graviton field would act {\it effectively} as a source of matter which drives the evolution of the spacetime.

The purpose of this paper is to enlighten the modifications that this new energy content brings to the curvature structure of the lower-dimensional black holes, and how it affects the maximum amount of energy that can be extracted from them.

Adopting the system of coordinates ($-\infty<t<+\infty$, $0<z<+\infty$, $-\infty<x<+\infty$), and the Lorentzian signature $[-,+,+]$, one of the simplest solutions of the above theory would be a static and spherically symmetric massive BTZ black hole metric with the following form \cite{Hendi:2016pvx}
\begin{eqnarray}
\label{metric}
\d s^2 &=& \frac{1}{z^2}  \left[-f(z) \d t^2 + \frac{\d z^2} {f(z)} +  \d x^2  \right], \\
f(z)&=& -\Lambda-mz^2-2q^2z^2 \ln \left(  \frac{1}{zl}\right) +M^2 z\,,
\end{eqnarray}
in which $m$ and $q$ are the total mass and the electric charge of the black hole respectively, and $\Lambda$ is the cosmological constant. The parameter $M$ denotes the mass of graviton field here. Also, $l$ is a reference length scale needed for having a dimensionless term in the argument of the logarithm and it could be set to 1 without any loss of generality, since it could be re-absorbed into $z$ through the rescaling of the coordinates.

The location of the horizon is given implicitly by the condition $f(z_{\rm hor})=0$ \cite{Hendi:2016pvx}. In the case of a black hole without electric charge, the following analytical solution could be found
\begin{eqnarray}
z_{\rm hor}&=& \frac{M^2}{m},  \qquad {\rm if } \qquad \Lambda=0, \\
z_{\rm hor}&=& \sqrt{-\frac{\Lambda}{m}},  \qquad {\rm if } \qquad M=0, \\
z_{\rm hor}&=& \frac{M^2 \pm \sqrt{M^4- 4m\Lambda}}{2m},  \qquad {\rm if } \qquad M, \ \Lambda \ne 0.
\end{eqnarray}

As expected, more massive gravitons would make the size of the black hole shrink. This intuitive behavior is a posteriori confirmation of the choice of $\mathcal{U}_{1}=c_0/r$.  

In fact, nearby the black hole, the dominant term within the metric factor comes from its own mass, and the electric charge would have the second-order effect, while asymptotically, the spacetime behavior is mostly affected by only the cosmological constant, with the graviton mass being the dominant effect in the intermediate regions. 

\begin{figure}
	\begin{center}
    $
    \begin{array}{cc}
{\includegraphics[scale=0.45, angle=0]{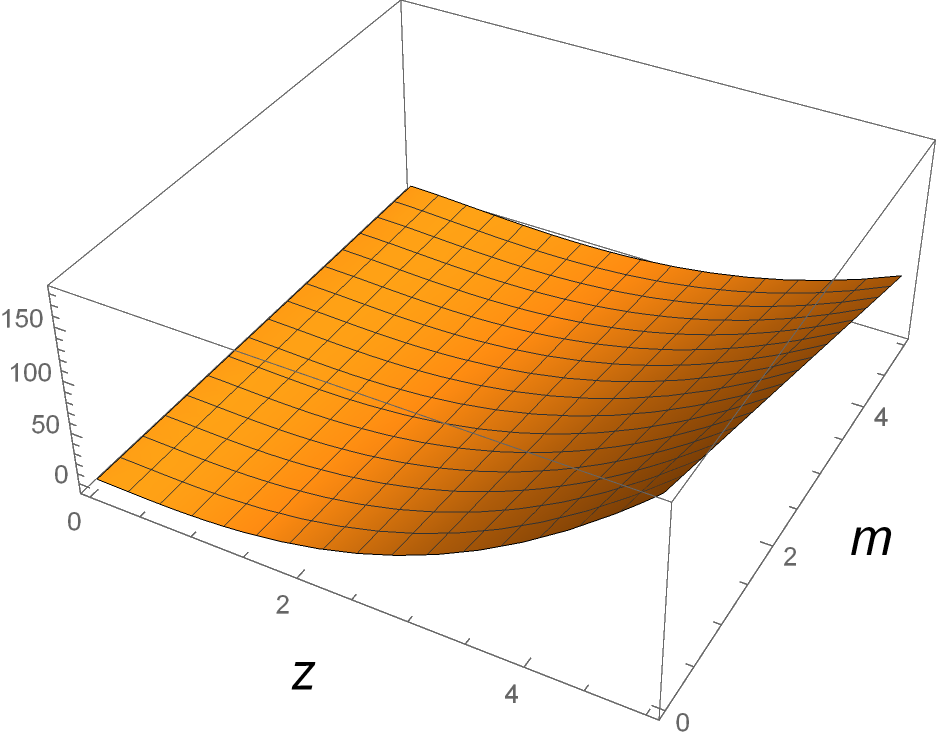}}
\hspace{3mm}
{\includegraphics[scale=0.45, angle=0]{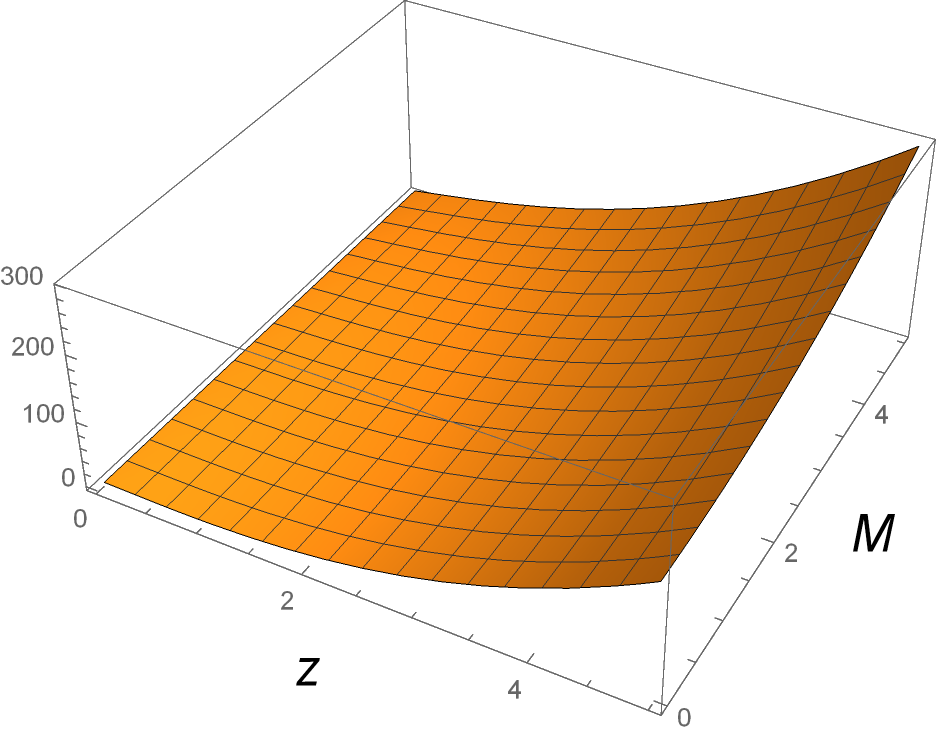}}\\
(a) \hspace{40mm} (b)  \\[7mm]
{\includegraphics[scale=0.45]{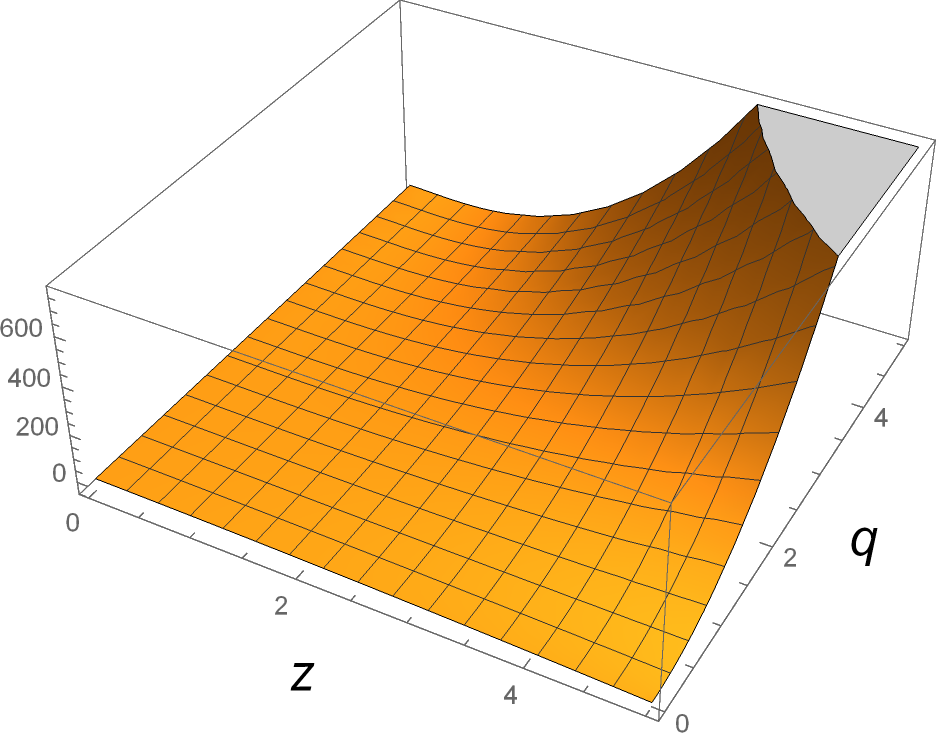}}
\hspace{3mm}
{\includegraphics[scale=0.45]{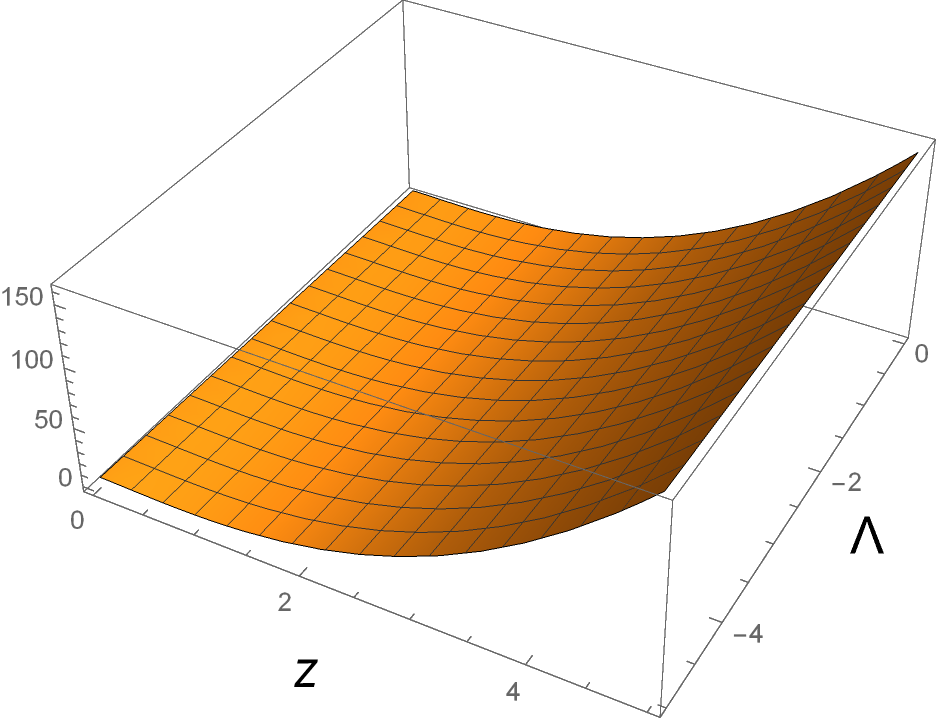}}\\
 (c) \hspace{40mm} (d) 
\end{array}
$
\end{center}
\caption{The figure displays the behavior of the metric function $f(z)$ vs. the radial coordinate for the following situations: in panel (a) the parameters are fixed as $\Lambda= -1.0$, $M = 0.5$, $q = 1.5$,  in panel (b)  they are $\Lambda= -1.0$, $m = 0.5$, $q = 1.5$, in panel (c), $\Lambda= -1.0$, $M = 0.5$, $m = 1.5$,  and in panel (d) we set $m= 1.0$, $M = 0.5$, $q = 1.5$. The plots show that the scalar polynomial curvature is an increasing function of the radial coordinate $z$, and an almost-constant function on this length scale with respect to the parameters of the solution $M$, $m$ and $\Lambda$, but the electric charge $q$ would increase it.}
\label{figf}
\end{figure}

Our analysis could be confirmed by the results of the numerical investigations shown in Fig. (\ref{figf}). In panel (a) the parameters are fixed as $\Lambda= -1.0$, $M = 0.5$, $q = 1.5$,  in panel (b)  they are $\Lambda= -1.0$, $m = 0.5$, $q = 1.5$, in panel (c), we have $\Lambda= -1.0$, $M = 0.5$, $m = 1.5$, and finally in panel (d) we set $m= 1.0$, $M = 0.5$, $q = 1.5$.

By an explicit computation we obtain the following relation between the parameters, 
\begin{eqnarray}
\label{poly}
{\mathcal I}_1 \,:=\,  R_{abcd;e}  R^{abcd;e} &\,=\,&   z^2 f(z)\Big [  z^4 (f'''(z))^2 +4 (z f''(z)-f'(z))^2 \Big]\\
&\,=\,& 4z^2 f(z)(20 q^4 z^2-8 M^2 q^2 z+M^4)\,. \nonumber
\end{eqnarray}

This account a polynomial scalar curvature invariant. The semicolon here denotes a covariant derivative and the prime denotes derivative with respect to $z$.

This curvature quantity would vanish on the horizon (since it vanishes for the case of $f(z)=0$). Moreover, the root of $z=0$ does not constitute a false positive, since this point does not actually belong to the spacetime manifold.

 Then, one could note that the discriminant of the quadratic equation $20 q^4 z^2-8 M^2 q^2 z+M^4=0$ would  read as
\beq
\Delta=-16 (Mq)^4<0\,,
\eeq
and therefore there are no other roots for the algebraic equation of ${\mathcal I}_1  =0$. 

Finally,  looking at the signs, one could get the relation sgn[${\mathcal I}_1$]=sgn[$f(z)$], and therefore the quantity ${\mathcal I}_1$ switches its sign, with a change of the sign of the metric component, (which actually would correspond to a change in the Lorentzian signature of the metric between $t$ and $z$). 

\begin{figure}
\begin{center}
    $
    \begin{array}{cc}
{\includegraphics[scale=0.45, angle=0]{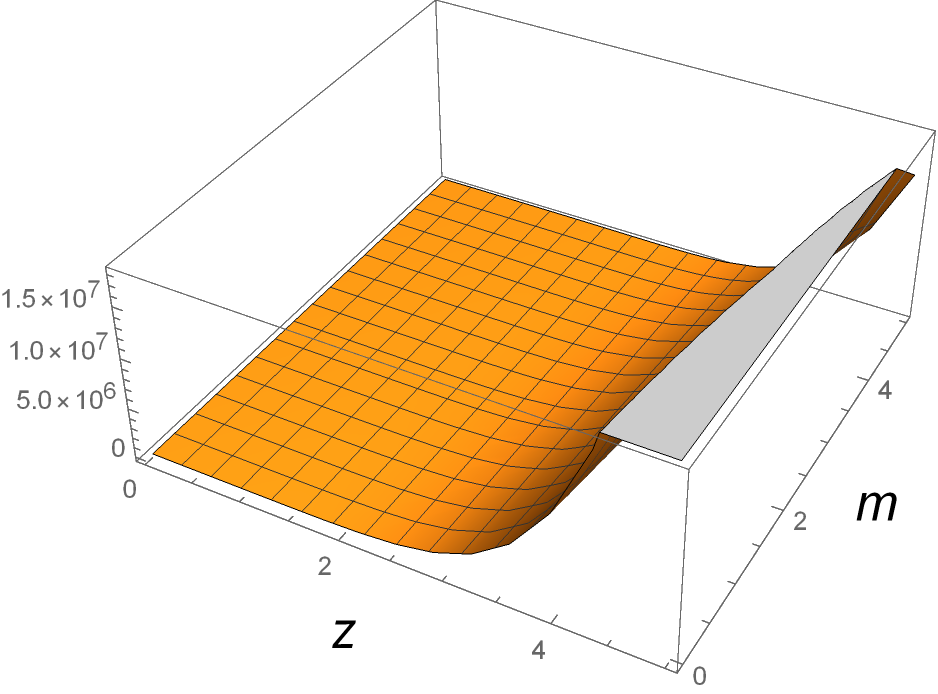}}
\hspace{3mm}
{\includegraphics[scale=0.45, angle=0]{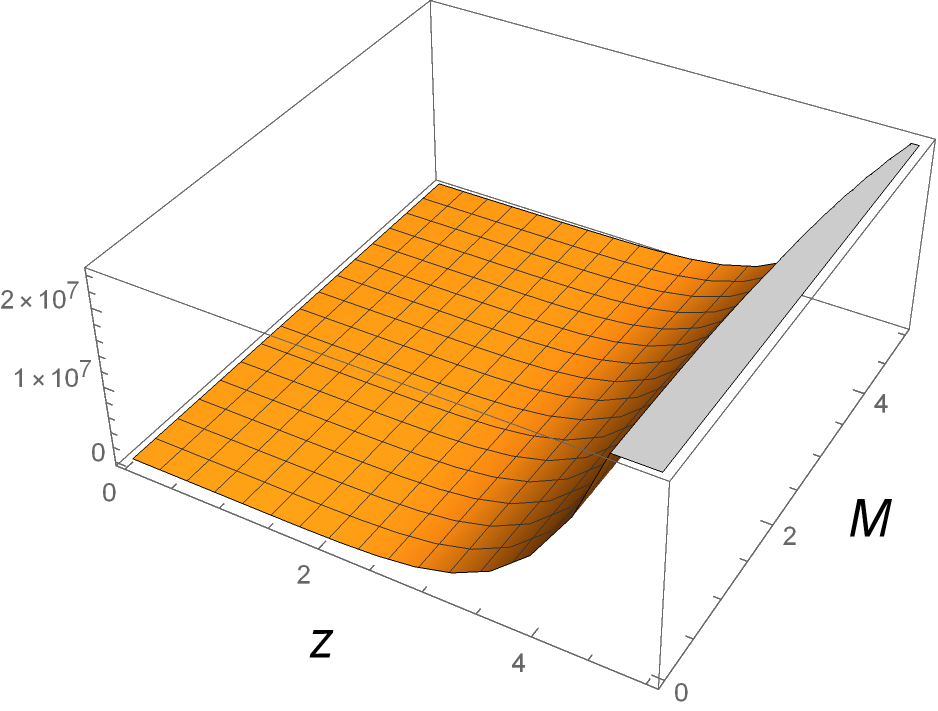}}\\
(a) \hspace{40mm} (b)  \\[7mm]
{\includegraphics[scale=0.45]{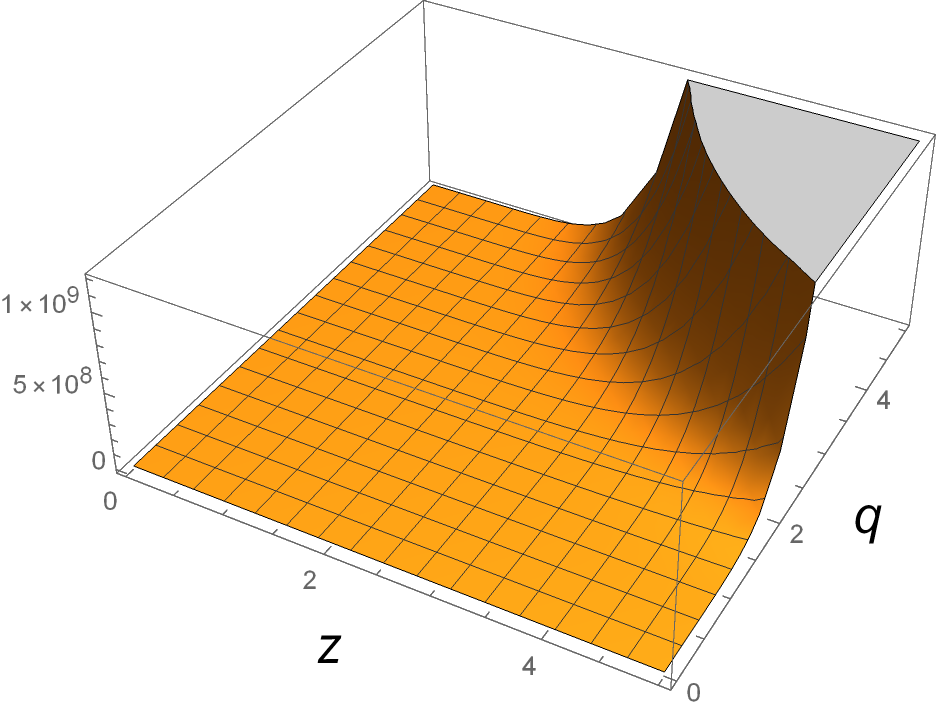}}
\hspace{3mm}
{\includegraphics[scale=0.45]{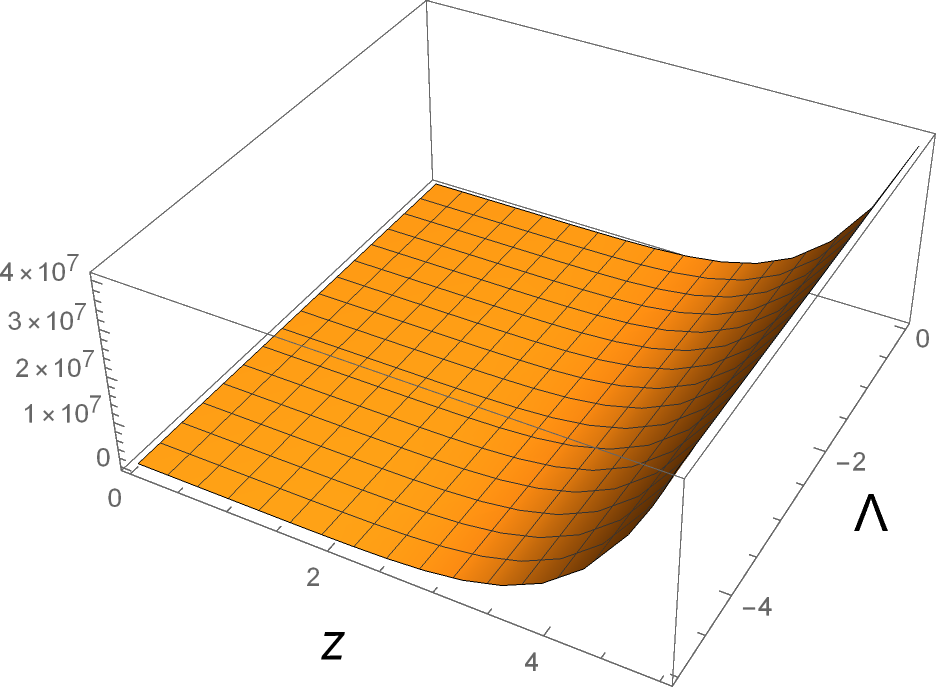}}\\
 (c) \hspace{40mm} (d) 
\end{array}
$
\end{center}
\caption{In panel (a) the parameters are fixed as $\Lambda= -1.0$, $M = 0.5$, $q = 1.5$,  in panel (b) they are $\Lambda= -1.0$, $m = 0.5$, $q = 1.5$, in panel (c), the parameters are $\Lambda= -1.0$, $M = 0.5$, $m = 1.5$ (we note that for this choice of numerical values of the parameters we get ${\mathcal I}_1(q=0)=0.25 z^2 (1. + 0.25 z - 1.5 z^2)\neq0$) and in panel (d) we set $m= 1.0$, $M = 0.5$, $q = 1.5$. The plots show that the scalar polynomial curvature is an increasing function of the radial coordinate $z$, and an almost constant function with respect to the parameters of the solution $M$, $m$ and $\Lambda$, but the electric charge $q$ would increase it.}
\label{figa}
\end{figure}

In Fig. (\ref{figa}) we display the radial behavior of the scalar polynomial curvature invariant ${\mathcal I}_1$ as a function of the coordinate $z$. The values of the parameters are the same as those in Fig. (\ref{figf}), and we could note that for this specific choice of numerical values for the parameters, we would get the relation ${\mathcal I}_1(q=0)=0.25 z^2 (1. + 0.25 z - 1.5 z^2)\neq0$.

The plots show that the scalar polynomial curvature is an increasing function of the radial coordinate $z$, and an almost-constant function with respect to the parameters of the solution $M$, $m$ and $\Lambda$, but the electric charge $q$ would increase it. 

Furthermore, the mass of the graviton acts as a scaling effect with respect to the evolution of the function $f(z)$ exhibited in Fig. (\ref{figf}). 

\begin{figure}
\begin{center}
    $
    \begin{array}{cc}
{\includegraphics[scale=0.45, angle=0]{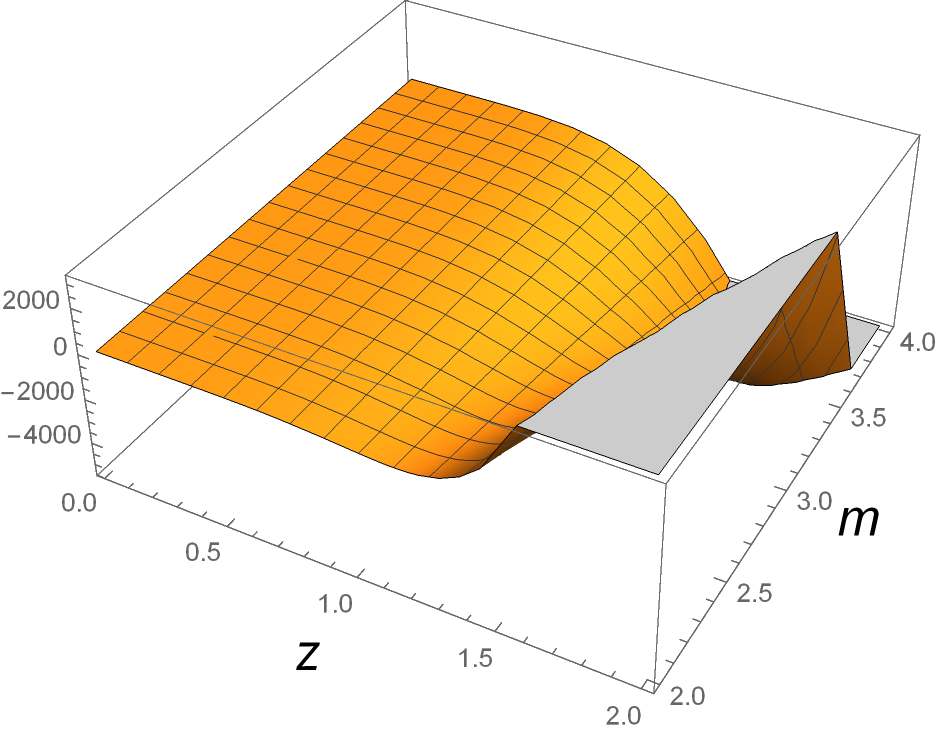}}
\hspace{3mm}
{\includegraphics[scale=0.45, angle=0]{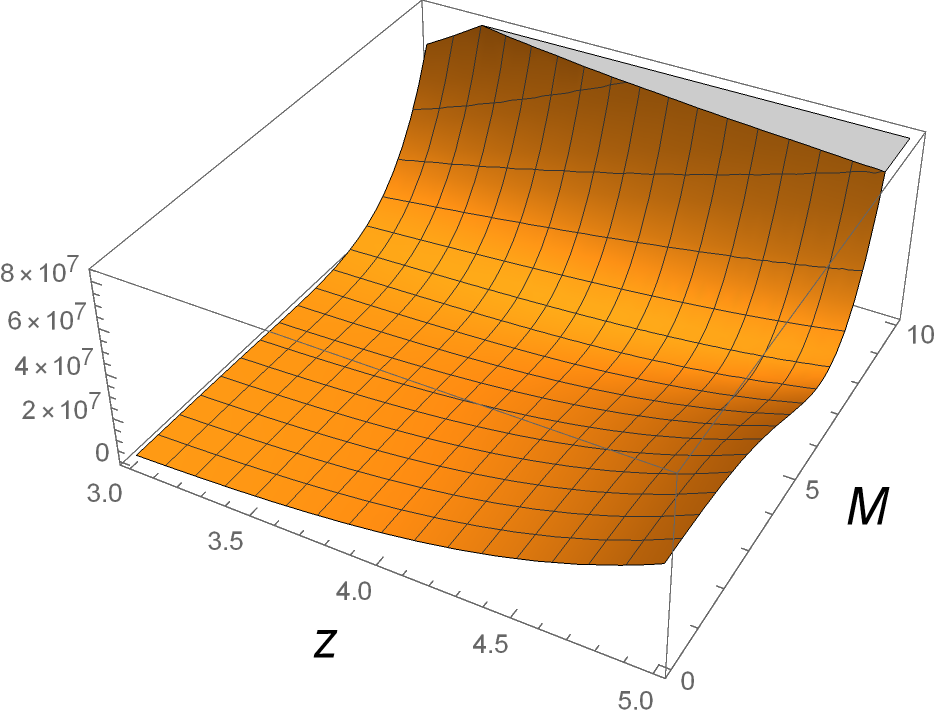}}\\
(a) \hspace{40mm} (b)  \\[7mm]
{\includegraphics[scale=0.45]{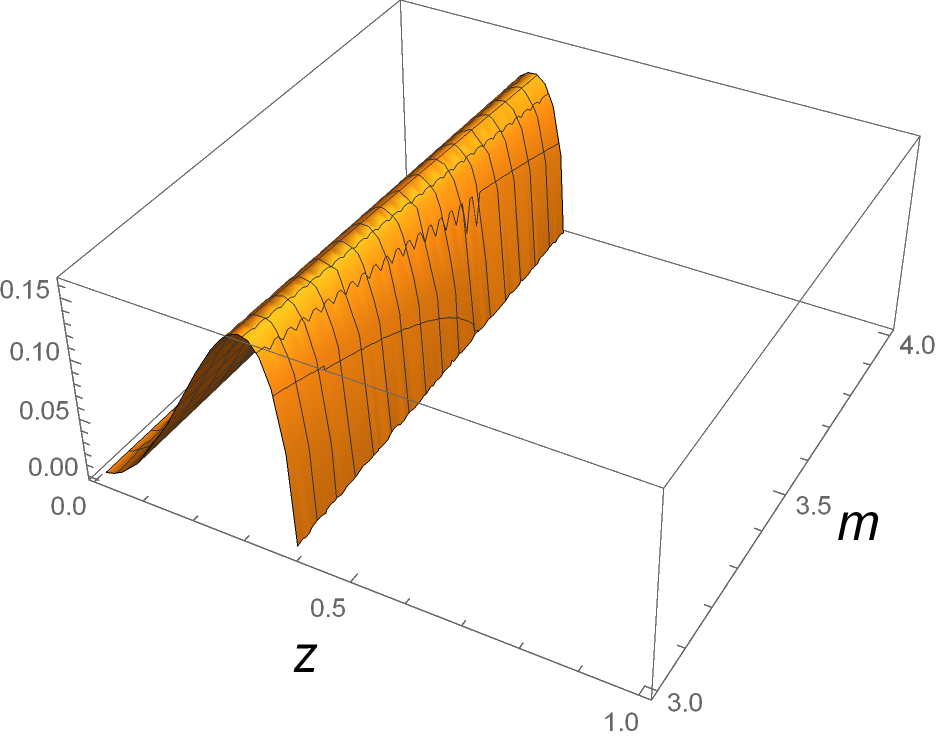}}
\hspace{3mm}
{\includegraphics[scale=0.45]{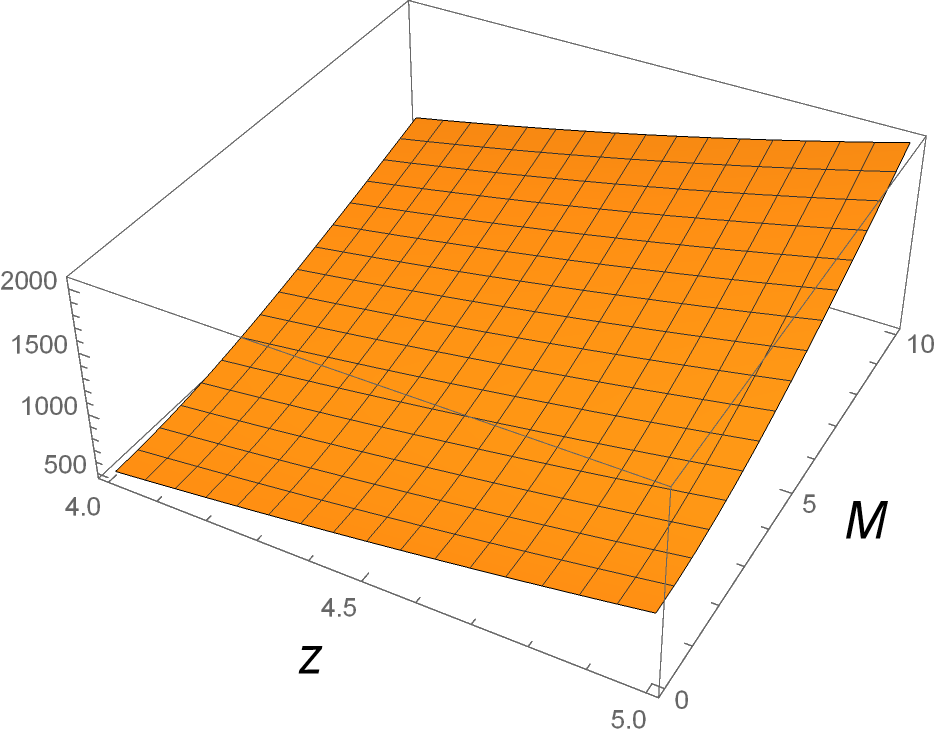}}\\
 (c) \hspace{40mm} (d) 
\end{array}
$
\end{center}
\caption{The figure displays the radial behavior of the scalar polynomial curvature invariant in panels (a)-(b) and of the Cartan curvature invariant in panel (c)-(d). They have been found by zooming on specific ranges of the radial distance starting from the event horizon of black hole. We note that a qualitative difference between panels (a) vs. (c) is that in the latter case the Cartan curvature invariant could not any longer be defined after reaching its minimum. Note that in panels (a)-(c) we have fixed $\Lambda= -1.0$, $M = 0.5$, $q = 1.5$ and in panels (b)-(d) we have fixed $\Lambda= -1.0$, $m = 0.5$, $q = 1.5$. 
}
\label{focus}
\end{figure}

However, a closer zoom at specific distances from the black hole location and restricting the ranges of the values of the free parameters does indeed would point out to a non-trivial structure with respect to the black hole parameters, as shown in panels (a)-(b) of Fig. (\ref{focus}).

\begin{figure}
\begin{center}
    $
    \begin{array}{cc}
{\includegraphics[scale=0.45, angle=0]{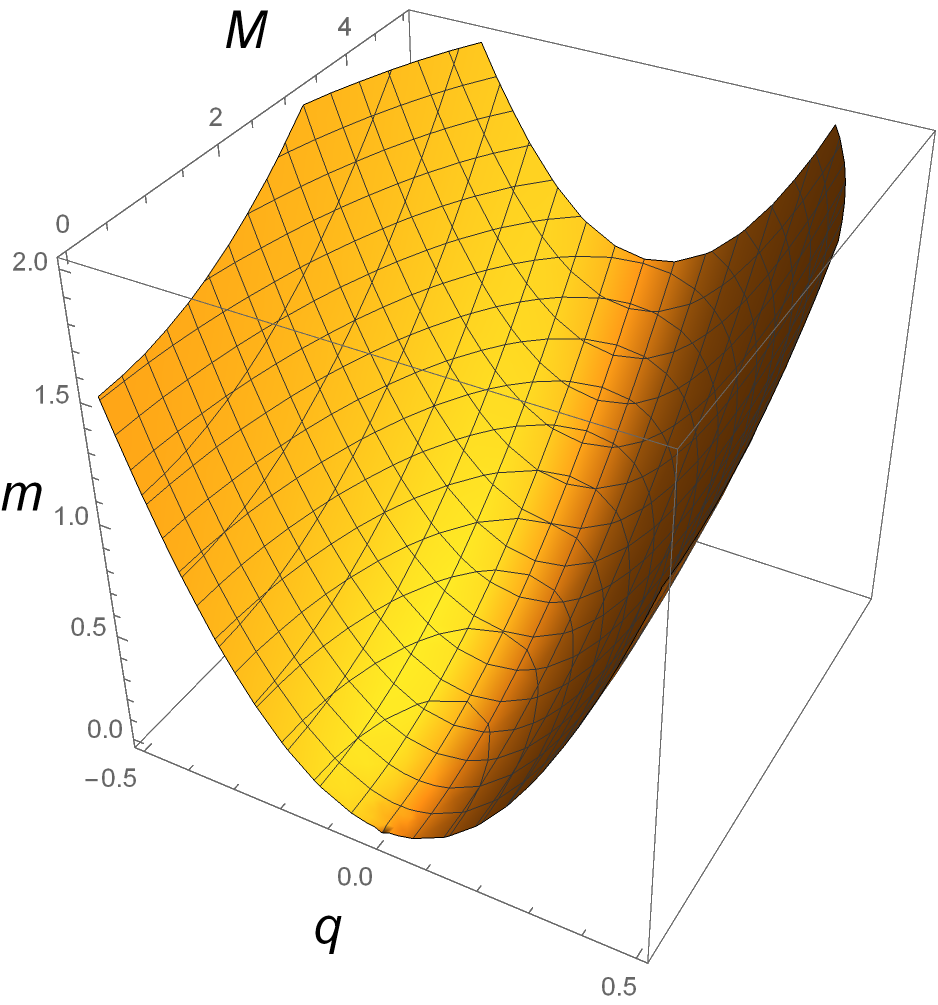}}
\hspace{3mm}
{\includegraphics[scale=0.45, angle=0]{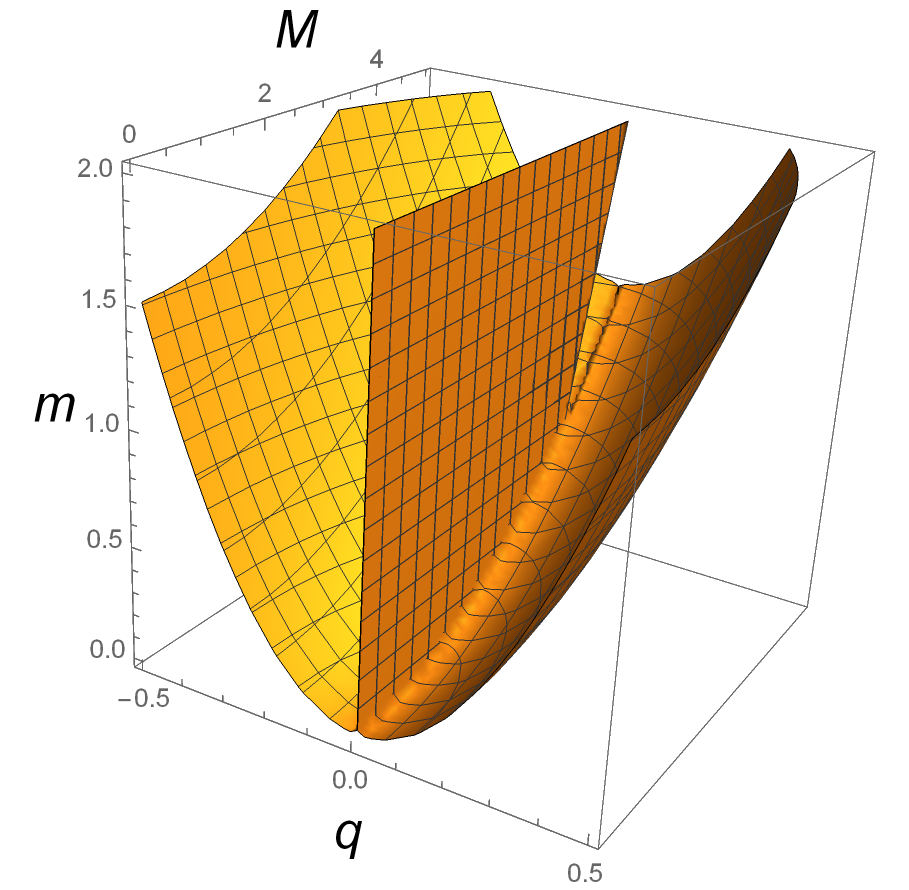}}\\
(a) \hspace{40mm} (b)  
\end{array}
$
\end{center}
\caption{The figure displays the level surface for ${\mathcal I}_1=0$  in panel (a), and for ${\mathcal J}^2_1=0$ in panel (b). We have set $z=20$ and $\Lambda=-5$. This comparison enlightens the fact that  ${\mathcal I}_1$  is a well-behaving quantity for locating the black hole horizon even for an electrically neutral black hole, contrary to ${\mathcal J}_1$. This results is a consequence of a non-zero mass for the graviton field.  In the case of massless gravity neither ${\mathcal I}_1$  nor ${\mathcal J}_1$ could locate the black hole horizon. 
}
\label{figc}
\end{figure}

In Fig. (\ref{figc}), panel (a), we also present the equation ${\mathcal I}_1=0$ as a level surface for $z=20$ and $\Lambda=-5$.

Then, the Cartan curvature invariant which detects the horizon could be found as
\beq
\label{cartan}
{\mathcal J}_1= {\bf e}_z (R_{ztzt}) \,=\, \frac{ z^3 \sqrt{2 f(z)}  f'''(z)}{4}   \,=\,  (qz)^2 \sqrt{2 f(z)}   \,,
\eeq
where ${\bf e}_z:= \sqrt{g^{zz}}\partial_z= z\sqrt{f(z)} \partial_z$  denotes a frame derivative along the $z$ direction.

We computed this quantity with respect to the coframe of 
\begin{flalign}
\label{triad}
 {l_a}&=\frac{1}{z} \left(\sqrt{\frac{ f(z )}{2}} \d t-\sqrt{\frac{1}{2f(z)}} \d z\right)\,, \nonumber\\ \qquad  {n_a}&=\frac{1}{z} \left(\sqrt{\frac{ f(z )}{2}} \d t+\sqrt{\frac{1}{2f(z)}} \d z\right)\,, \nonumber\\ \qquad  {m_a}&=\frac{1}{\sqrt{2} z} \d x \,,
\end{flalign}
where the metric (\ref{metric}) reads as
\beq
\label{metricframe}
g_{ab}= -2 l_{(a} n_{b)} +2 m_{(a} m_{b)}= {\bf e}_a \cdot {\bf e}_b \,.
\eeq

In the above relation, the round parentheses denote symmetrization.

The ``null" triad (\ref{triad}) satisfies the following relations
\beq
\label{tr3}
m_a m^a =1 =- l_a n^a \,, \qquad l_a l^a = n_a n^a=l_a m^a=n_a m^a=0 \,,
\eeq
where they constitute the appropriate canonical coframe for computing the Cartan invariant.

Computing the following five Ricci scalars
\begin{eqnarray}
\Phi_{00}&=&\frac{1}{2}R_{ab} l^a l^b\,, \\
\Phi_{22}&=&\frac{1}{2}R_{ab} n^a n^b \,, \\
\Phi_{10}&=&\frac{1}{2\sqrt{2}}R_{ab} m^a l^b \,, \\
\Phi_{12}&=&\frac{1}{2 \sqrt{2}}R_{ab} m^a n^b \,, \\
\Phi_{11}&=&\frac{1}{6}  (R_{ab} m^a m^b +R_{ab} n^a l^b)\,,
\end{eqnarray}
we note that all of them except $\Phi_{11}$ would vanish.

Therefore, we could reduce the curvature to its canonical form since the symmetry group of our metric is $SO(1,1)$, as our black hole solution is a static and spherically symmetric metric \cite{Sousa:2007ax}. 

Note that this procedure has removed all of the possible ambiguities in the construction of the coframe which could be rescaled under boosts \cite{Ashtekar:2002qc},
\beq
{l_a} \to C_1  {l_a}, \qquad {n_a} \to \frac{1}{C_1} {n_a}, \qquad  {m_a} \to  {m_a},
\eeq
and under null rotations \cite{Ashtekar:2002qc},
\beq
{l_a} \to \frac{1}{2} {l_a}, \qquad  {n_a} \to \frac{1}{2}C_2^2  {l_a}+\frac{1}{2} {n_a}+C_2 {m_a}, \qquad  {m_a} \to C_2  {l_a} + {m_a}\,,
\eeq
where in the above relations, $C_1=C_1(t,z,x)$ and $C_2=C_2(t,z,x)$ are two arbitrary functions of the manifold coordinates. 

These Lorentz transformations leave the spacetime metric unchanged, but they could affect the components of the curvature.

One could then explicitly derive the following relation
\beq
\label{phi11}
\Phi_{11}=\frac{1}{3}\left( -M^2 z +4\Lambda -\frac{q^2}{2 z^2}+\frac{M^2}{4z^3}-\frac{\Lambda}{2z^2} \right)\,.
\eeq

We could see that the mass of the gravitons has the leading effect in shaping this curvature invariant, both in the far-field and in the local-field limits. 

It is worth mentioning that the massive gravity theory would treat the mass of the graviton as a source of matter in the spacetime. This is due to the fact that this term would explicitly enter the field equations which have been derived from the Lagrangian (\ref{Action}), and also it could effectively be re-absorbed into its stress-energy tensor.  This is also true for the case of black hole electric charge in the coupled Maxwell-Einstein theory. 

On the other hand, the mass of the black hole does not enter the Ricci curvature scalar, which is also the same with the case of the well-known Schwarzschild solution.

\begin{figure}Ne
\begin{center}
    $
    \begin{array}{cc}
{\includegraphics[scale=0.45, angle=0]{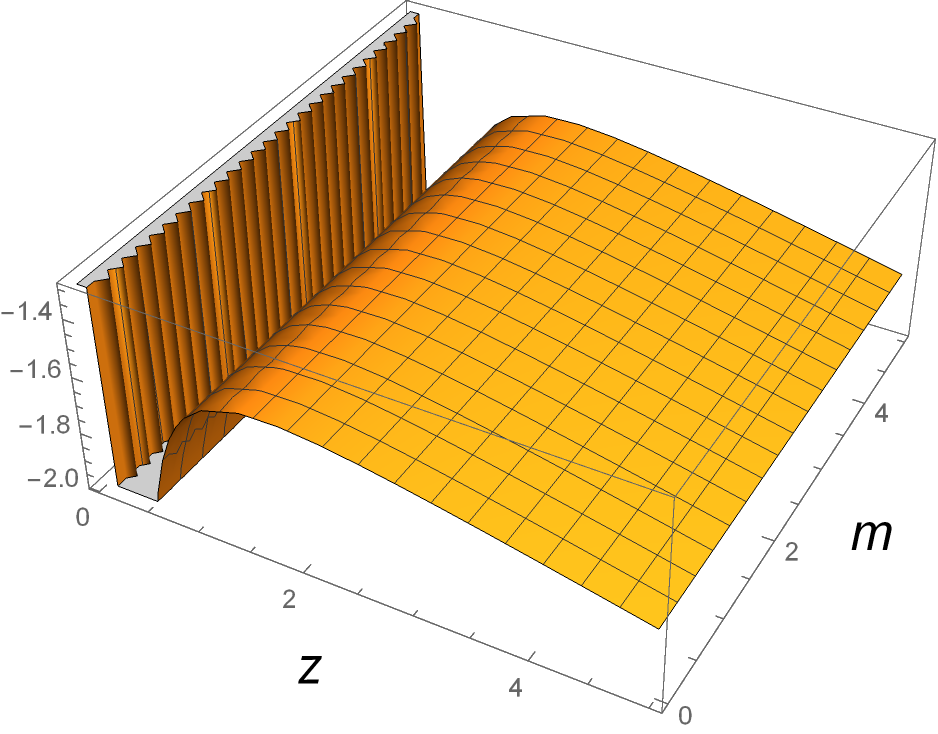}}
\hspace{3mm}
{\includegraphics[scale=0.45, angle=0]{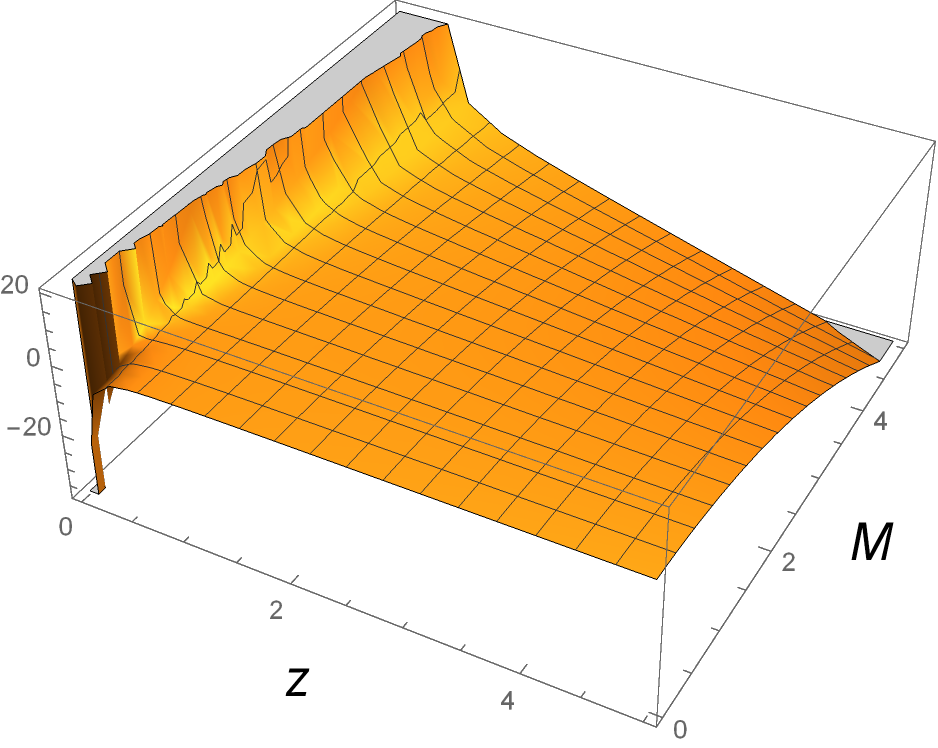}}\\
(a) \hspace{40mm} (b)  \\[7mm]
{\includegraphics[scale=0.45]{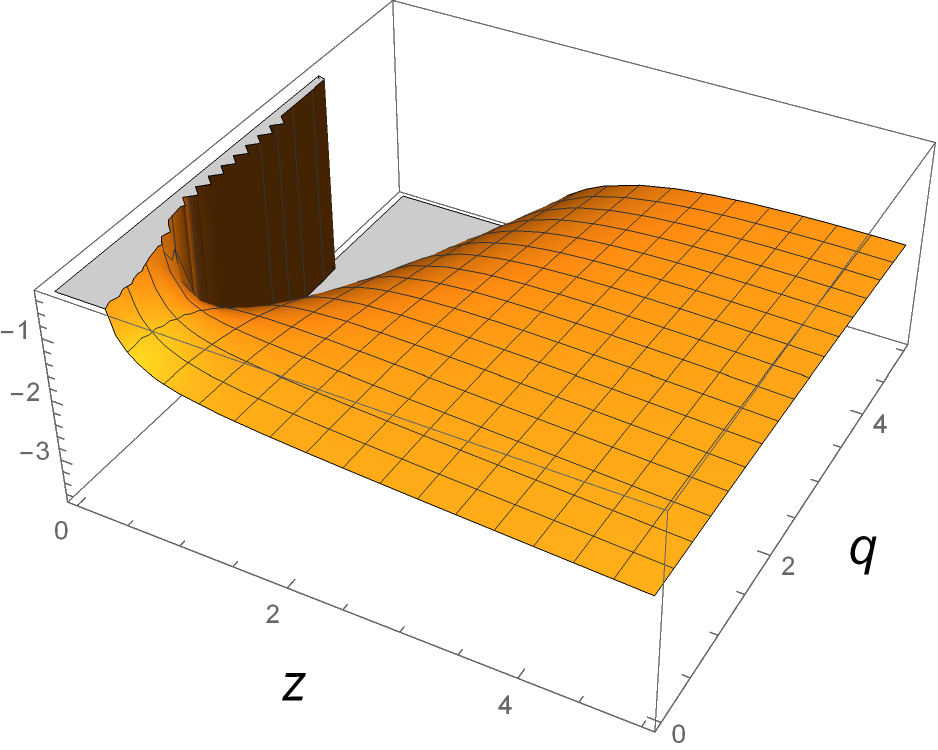}}
\hspace{3mm}
{\includegraphics[scale=0.45]{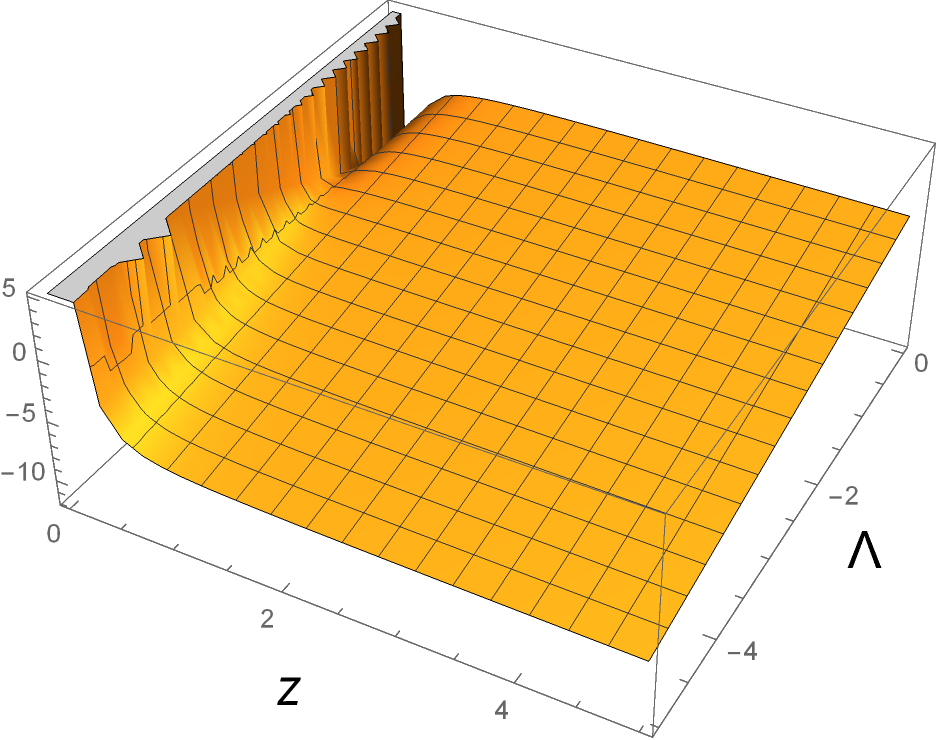}}\\
 (c) \hspace{40mm} (d) 
\end{array}
$
\end{center}
\caption{The figure displays the behavior of the Ricci curvature scalar $\Phi_{11}$ vs. the radial coordinate $z$ in the following situations: in panel (a) the parameters are fixed as $\Lambda= -1.0$, $M = 0.5$, $q = 1.5$,  in panel (b)  they are $\Lambda= -1.0$, $m = 0.5$, $q = 1.5$, in panel (c), $\Lambda= -1.0$, $M = 0.5$, $m = 1.5$  and in panel (d) we set $m= 1.0$, $M = 0.5$, $q = 1.5$. }
\label{figfi}
\end{figure}

In Fig. (\ref{figfi}) we show the numerical counterpart of this analytical discussion choosing the same values of the black hole parameters which we chose in Fig. (\ref{figf}).

One could also check that when all of the parameters of this black hole solution goes to zero, the remaining curvature quantity vanishes as well, which is in accordance with our expectation.

Following the previous discussions about the restrictions on the $z$ coordinates, one could also check that the Cartan invariant (\ref{cartan}) detects the horizon for us. We should emphasize that, there is actually only one functionally independent first order Cartan invariant\footnote{Two quantities $A=A(t,z,x)$ and $B=B(t,x,z)$ are said to be functionally dependent if $ ||dA \wedge d B||^2=0$, where $d$ is the exterior derivative.}. 

\begin{figure}
\begin{center}
    $
    \begin{array}{cc}
{\includegraphics[scale=0.45, angle=0]{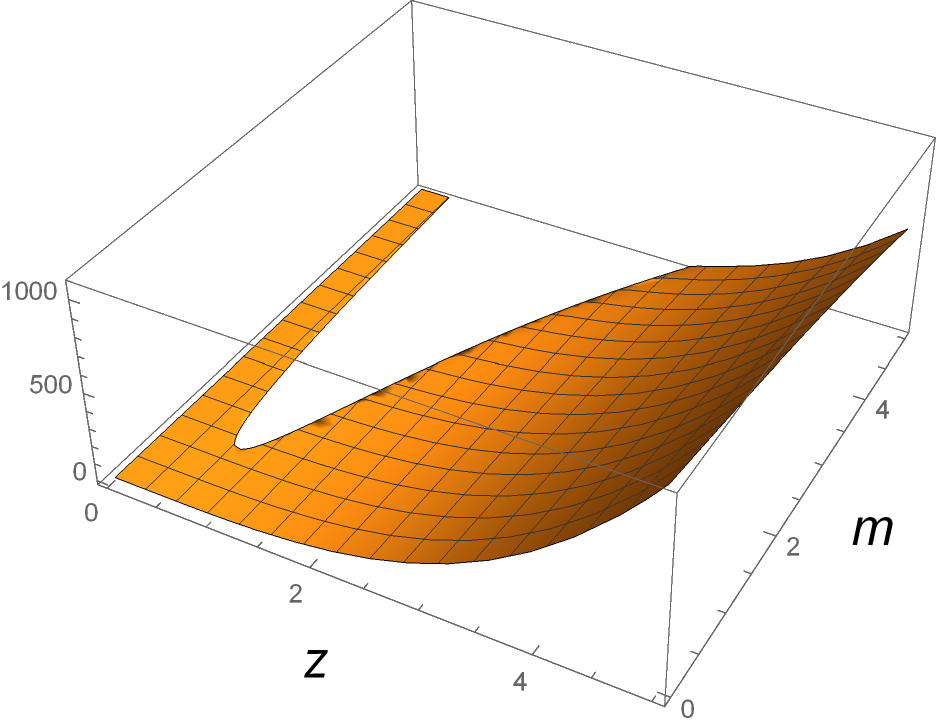}}
\hspace{3mm}
{\includegraphics[scale=0.45, angle=0]{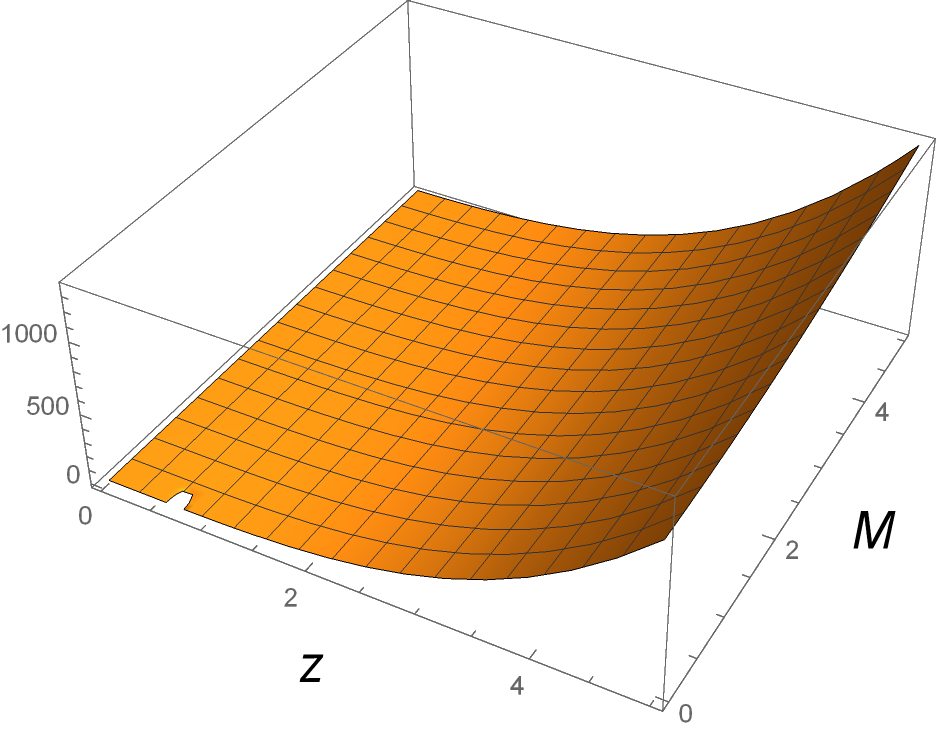}}\\
(a) \hspace{40mm} (b)  \\[7mm]
{\includegraphics[scale=0.45]{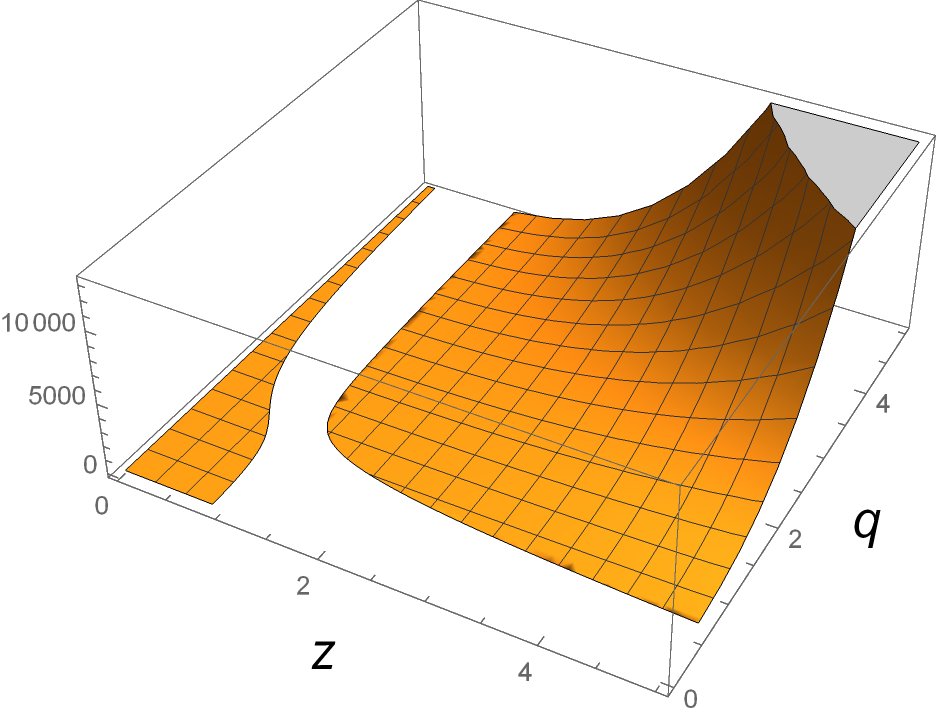}}
\hspace{3mm}
{\includegraphics[scale=0.45]{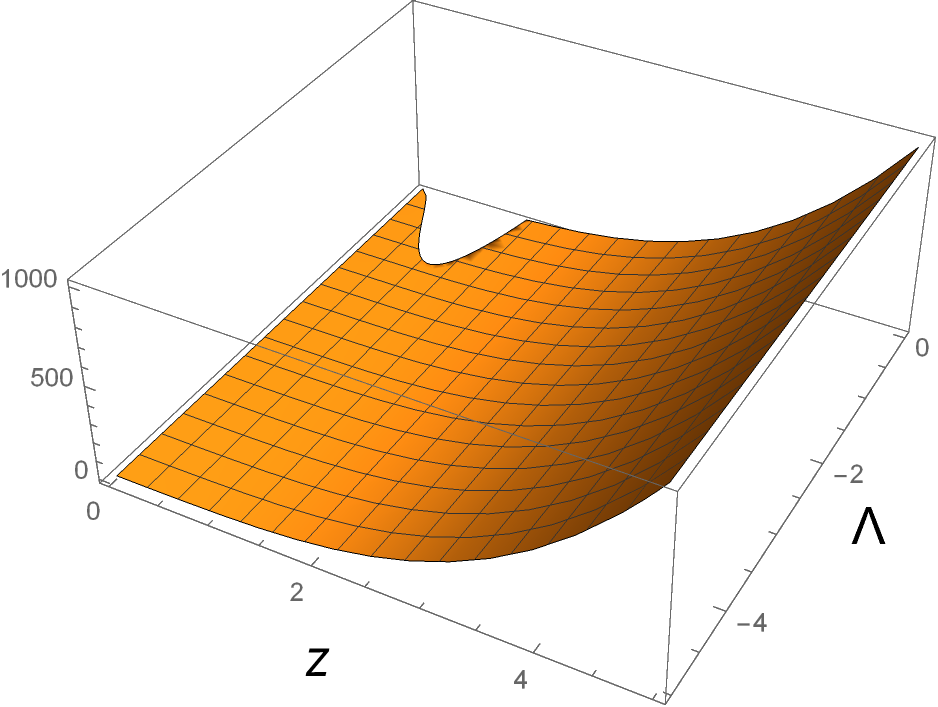}}\\
 (c) \hspace{40mm} (d) 
\end{array}
$
\end{center}
\caption{The figure displays the radial behavior of the Cartan curvature invariant ${\mathcal J}_1$. In panel (a) we have fixed $\Lambda= -1.0$, $M = 0.5$, $q = 1.5$;  in panel (b) we have fixed $\Lambda= -1.0$, $m = 0.5$, $q = 1.5$; in panel (c) we have fixed $\Lambda= -1.0$, $M = 0.5$, $m = 1.5$ and in panel (d) we have fixed $m= 1.0$, $M = 0.5$, $q = 1.5$. The plot shows that the scalar polynomial curvature is an increasing function of the radial coordinate $z$, and an almost-constant function with respect to the black hole parameters on this length scale (but for the case of the electric charge $q$). We note that while the sign of the metric function $f(z)$ does not affect the properties of the massive black hole solution and the applicability of the algorithm which relies on the scalar polynomial curvature invariant, it affects the Cartan method as it cannot be applied any longer for the choices of the black hole parameters which make $f(z)<0$.}
\label{figb}
\end{figure}

Here, in Fig. (\ref{figb}) we are displaying the radial behavior of the Cartan curvature invariant ${\mathcal{J}}_1$ as a function of the coordinate $z$. As for the parameters, in panel (a) we have fixed $\Lambda= -1.0$, $M = 0.5$, $q = 1.5$, in panel (b), $\Lambda= -1.0$, $m = 0.5$, $q = 1.5$, in panel (c), $\Lambda= -1.0$, $M = 0.5$, $m = 1.5$, and in panel (d) we set $m= 1.0$, $M = 0.5$, $q = 1.5$. 

These plots again show that the Cartan curvature invariant is an increasing function of the radial coordinate $z$, and it remains an almost-constant function with respect to the black hole parameters except for the case of the electric charge $q$. We also observe that the electric charge increases the Cartan curvature invariant.

Similar to the case of the scalar polynomial invariant, zooming from asymptote to a specific distance from the black hole location and restricting the ranges of the values of the free parameters could reveal the non-trivial structures, with respect to the black hole parameters, as displayed in panels (c)-(d) of Fig. (\ref{focus}). A key difference from the previous case, is that after reaching the minimum, the Cartan curvature invariant could not  be defined any longer, which is because of the change in the Lorentzian signature of the spacetime metric there.

We also note that the sign of the metric function $f(z)$ does not affect the properties of the massive black hole solution. This then preserves the applicability of the algorithm we employ here which actually relies on the scalar polynomial curvature invariant.  However, this sign could then affect the Cartan method, and so, for the choices of the black hole parameters for which $f(z)<0$, this method could not be applied any longer.

Note that the values of the black hole  parameters chosen in Figs. (\ref{figa}) and (\ref{figb}) are the same.

In Fig. (\ref{figc}), panel (b), we plot the equation ${\mathcal J}^2_1=0 $ as a level surface for $z=20$ and $\Lambda=-5$.\footnote{Note the zeroes of ${\mathcal J}^2_1$ and ${\mathcal J}_1$ are the same and we choose to plot the latter just for graphical convenience.}.

Our numerical results allow us to make further comments about the applicability of these two methods. First of all, the appropriate scalar polynomial curvature invariant is second degree in the curvature, while the appropriate Cartan invariant is first degree. This is the same result found in \cite{Gregoris:2019ycf} where the horizon of the BTZ black hole was studied in a massless gravity theory. However, a qualitative difference that we observe here is that, the Cartan method would fail for the vacuum solution where $q=0$, while the former method could still hold, even in that limit, which is in sharp contrary to the case of the massless gravity theory.

It would also worth to mention here that, in a very recent work \cite{Hendi:2020yah}, the geodesic trajectories of charged and massive BTZ black holes in these specific solutions of massive gravity have been presented, where the authors found that the electric charge parameter $q$ is in fact a critical parameter for classifying all the geodesic motions, for both of the time-like and also light-like particles. This is also what we have observed here, that the electric charge would greatly change the curvature invariants of our spacetime where the pattern could be comprehended further from the numerical studies.

\subsection{Massive gravity effects}

In this section, we try to deepen our investigations of the consequences of allowing a non-zero graviton mass.

\begin{figure}
\begin{center}
    $
    \begin{array}{cc}
{\includegraphics[scale=0.45, angle=0]{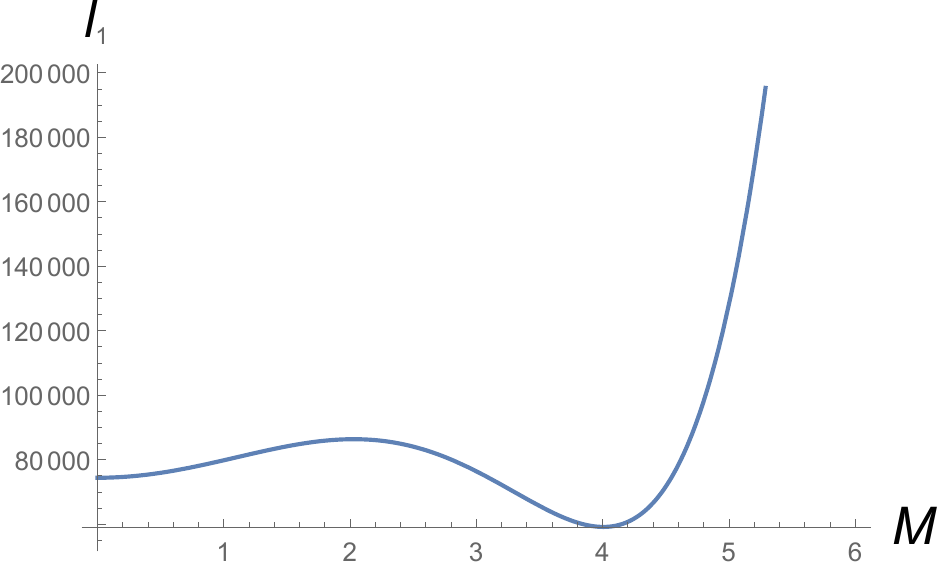}}
\hspace{3mm}
{\includegraphics[scale=0.45, angle=0]{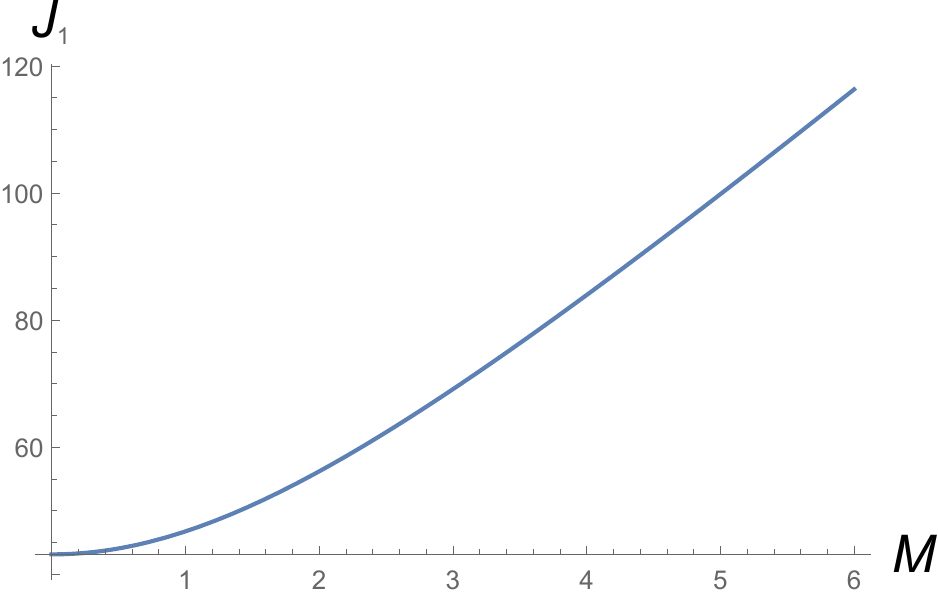}}\\
(a) \hspace{40mm} (b)  \\[7mm]
{\includegraphics[scale=0.45]{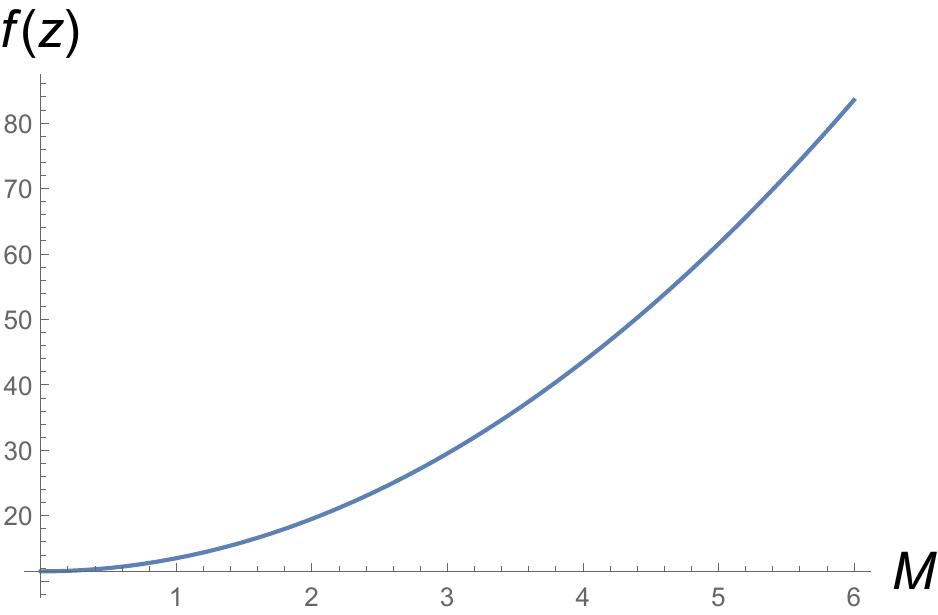}}
\hspace{3mm}
{\includegraphics[scale=0.45]{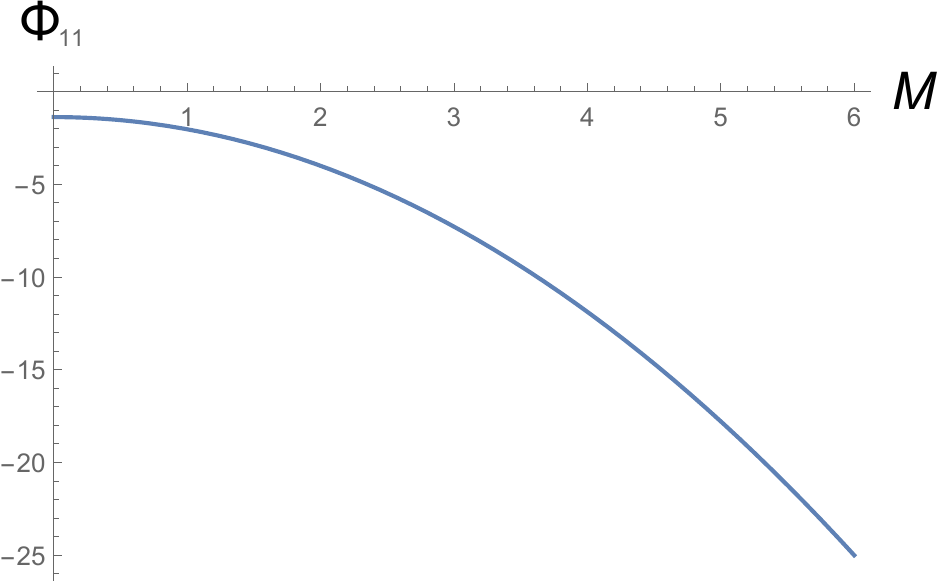}}\\
 (c) \hspace{40mm} (d) 
\end{array}
$
\end{center}
\caption{The figure  shows how the mass of the gravitons affects the polynomial scalar curvature invariant ${\mathcal I}_1$ in panel (a), of the Cartan invariant ${\mathcal J}_1$ in panel (b), of the metric function $f(z)$ in panel (c), and of the Ricci curvature $\Phi_{11}$ in panel (d) for the case ($z=2.0$, $\Lambda=-1.0$, $q=1.5$, $m=0.5$).
}
\label{figma}
\end{figure}

First, we present the results in Fig. (\ref{figma}), where in panel (a), we show how the mass of the gravitons is affecting the polynomial scalar curvature invariant ${\mathcal I}_1$, in panel (b) we show its effects on the Cartan invariant ${\mathcal J}_1$, in panel (c), we show the metric function $f(z)$, and in panel (d) we show the Ricci curvature $\Phi_{11}$. 

Note that these plots have been made by choosing the parameters as $z=2.0$, $\Lambda=-1.0$, $q=1.5$, $m=0.5$.  One could observe that while the evolution of the latter three quantities is monotonic with their absolute values increasing for more massive gravitons, the polynomial scalar curvature invariant exhibits a local maximum followed by a local minimum. 

\begin{figure}
\begin{center}
    $
    \begin{array}{cc}
    {\includegraphics[scale=0.4, angle=0]{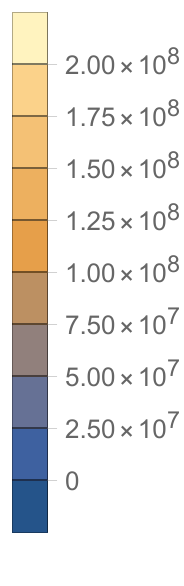}}
\hspace{0.5mm}
{\includegraphics[scale=0.4, angle=0]{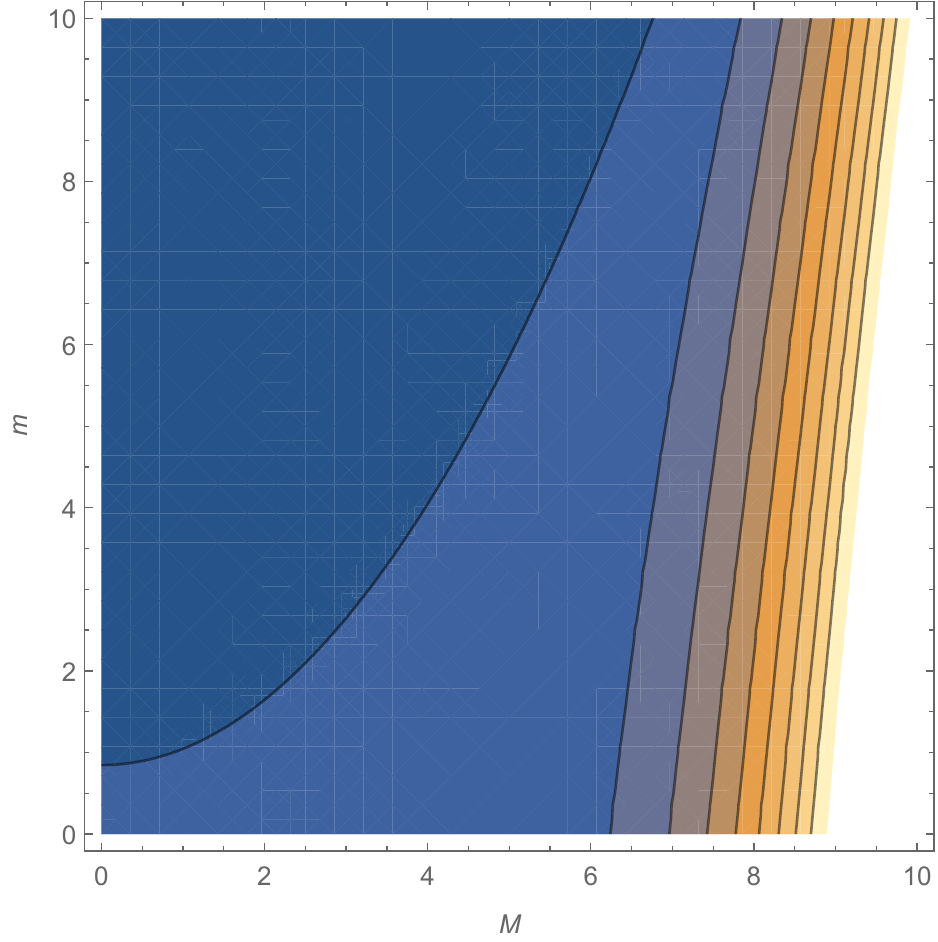}}
\hspace{3mm}
{\includegraphics[scale=0.4, angle=0] {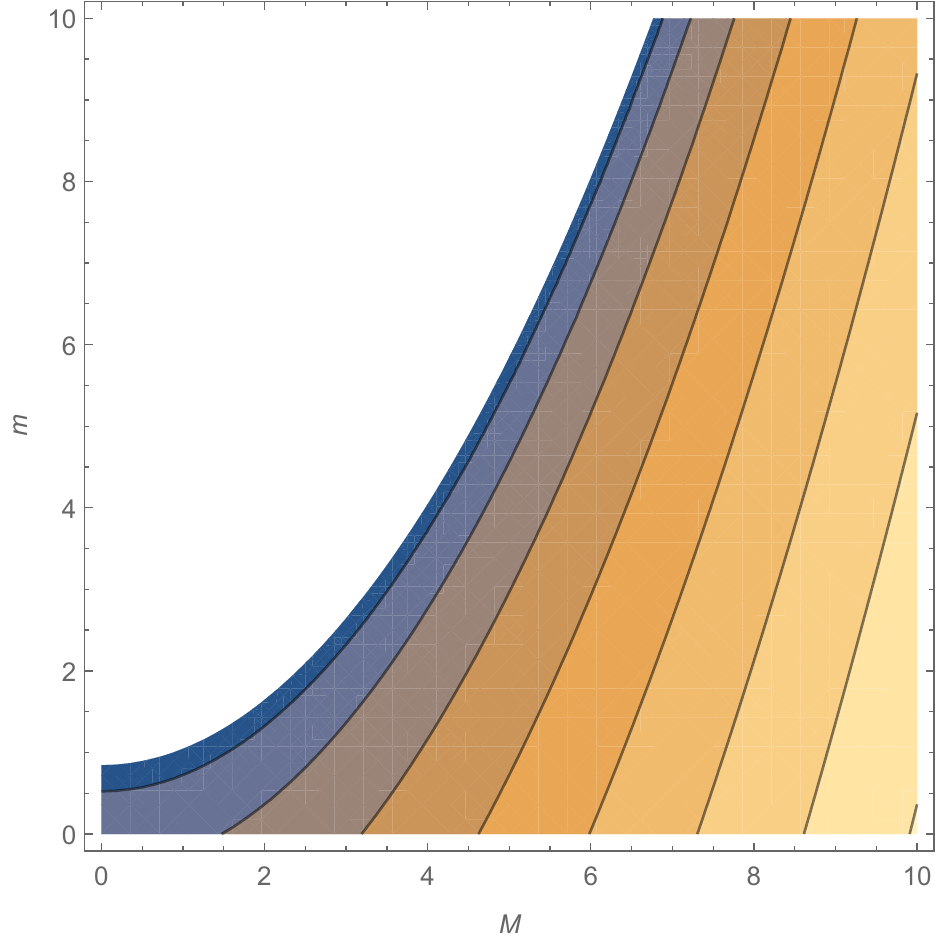}}  \hspace{0.5mm} {\includegraphics[scale=0.4, angle=0]{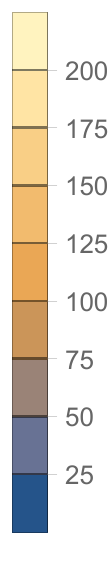}} \\
(a) \hspace{40mm} (b)  \\[7mm]
{\includegraphics[scale=0.4, angle=0]{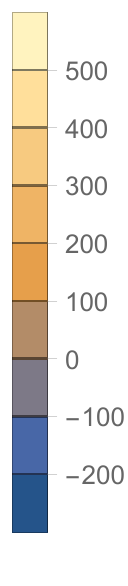}}
\hspace{0.5mm} {\includegraphics[scale=0.4]{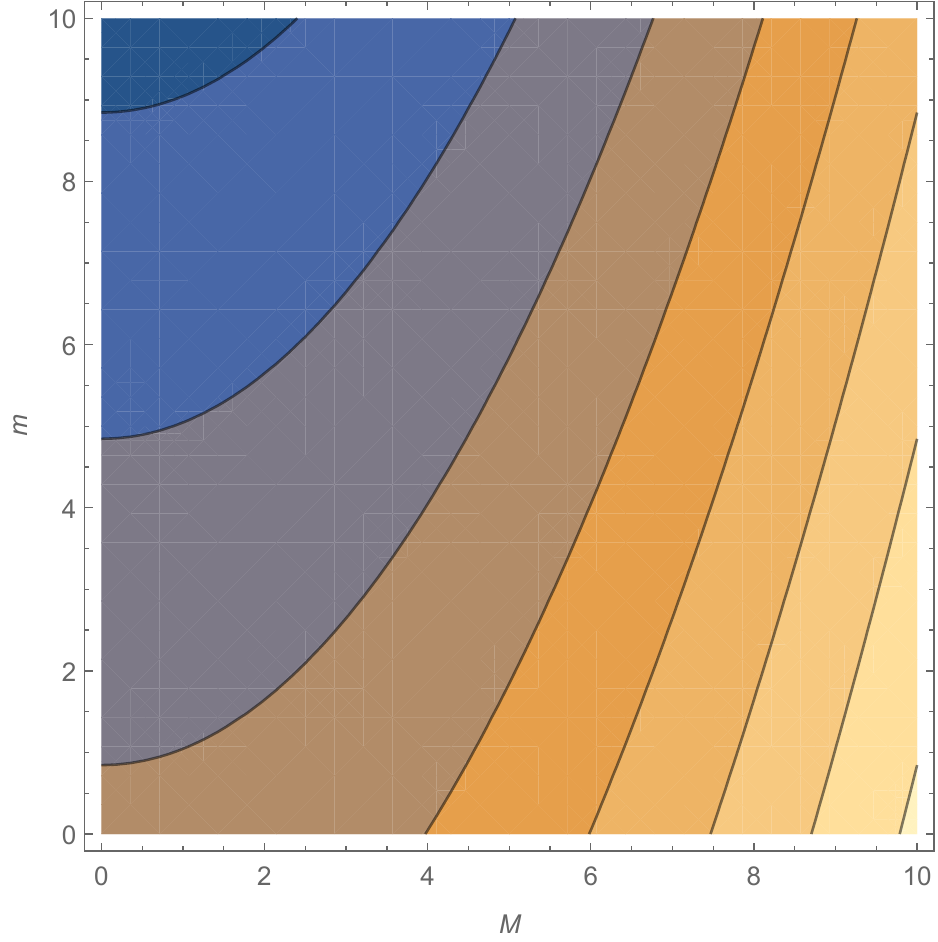}}
\hspace{3mm}
{\includegraphics[scale=0.4]{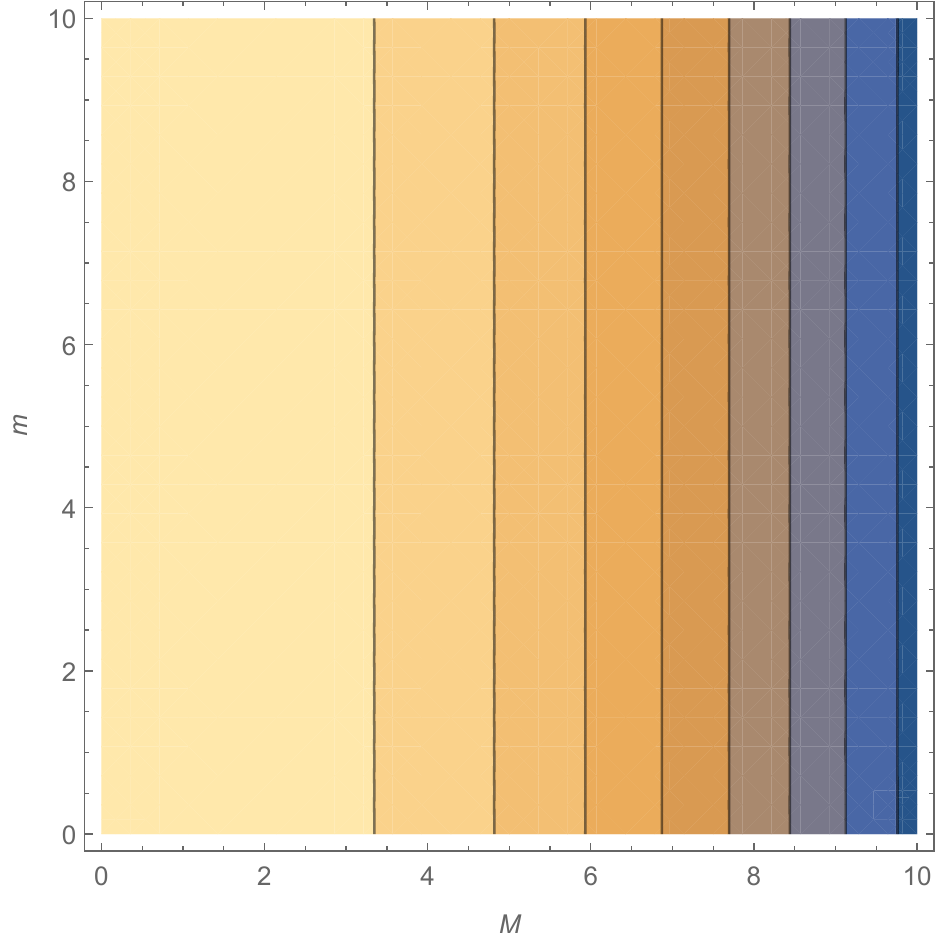}} \hspace{0.5mm} {\includegraphics[scale=0.4, angle=0]{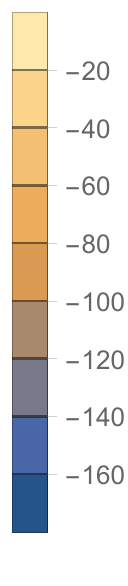}} \\
 (c) \hspace{40mm} (d) 
\end{array}
$
\end{center}
\caption{In the above figures,  in panel (a) we show the contourplots of ${\mathcal I}_1$ , in panel (b) the Cartan invariant ${\mathcal J}_1$ has been shown, in panel (c) the metric function $f(z)$ has been shown, and in panel (d), the Ricci curvature $\Phi_{11}$ is shown. These have been made choosing the parameters as $z=5.0$, $\Lambda=-1.0$, $q=0.5$. In particular, panel (b) is compatible with the discussion about Fig. (\ref{figb}) because when the spacetime switches its Lorentzian signature (i.e. when $f(z)$ changes sign), the Cartan curvature invariant will become ill-defined. Moreover, we confirm that the Ricci curvature invariant $\Phi_{11}$ is independent from the mass of the black hole, and depends linearly on the graviton mass, and that the "quadratic-hyperbola-law" for the location of the horizon $m z_{\rm hor} -M^2=0$ holds. 
}
\label{cont}
\end{figure}

 Next, in Fig. (\ref{cont}), in panel (a), we show the contourplots of ${\mathcal I}_1$, in panel (b), we show the Cartan invariant ${\mathcal J}_1$, in panel (c), we show the metric function $f(z)$, and finally in panel (d), we show the Ricci curvature $\Phi_{11}$. 
 
These second set of plots have been made by choosing the parameters as $z=5.0$, $\Lambda=-1.0$, $q=0.5$). Note that, in particular, our panel (c) is compatible with the discussions  below Fig. (\ref{figb}). So again we observe that when the spacetime switches its Lorentzian signature (i.e. when $f(z)$ changes sign), the Cartan curvature invariant would become ill-defined. 
 
Moreover, we could confirm here that the Ricci curvature invariant $\Phi_{11}$ is indeed independent of the mass of the black hole, and it depends linearly on the graviton mass. Also, from our results we could confirm the "quadratic-hyperbola-law", as  $m z_{\rm hor} -M^2=0$, which is for locating the horizon, would still hold here as well.

\subsection{Curvature structure of the BTZ black hole in massive gravity }

Now in this section, we study the structure of spacetime curvature of the metric (\ref{metric}) in more details. We also comment on the relationships between our results and those recent ones in the literature, which have investigated the relationships between various curvature quantities in (2+1)-dimensional spacetimes, such as \cite{Musoke:2015kql,Coley:2014goa}.

Here, we denote the Einstein and Ricci tensors with $G_{ab}$ and $R_{ab}$ respectively. Then, we could derive the following relations
\begin{eqnarray}
&&  R_{abcd}R^{abcd} - R_{ab}R^{ab}=\frac{z^2 (M^2-4q^2 z)^2}{2}, \\
&&  G_{ab;c} R^{ab;c}= 4 z^3 q^2 (4 q^2 z-M^2) f(z) \,.
\end{eqnarray}

From these, we then get
\beq
\frac{  R_{abcd;e}  R^{abcd;e}}{ R_{abcd}R^{abcd} - R_{ab}R^{ab}}=\frac{8   f(z)(20 q^4 z^2-8 M^2 q^2 z+M^4)}{(M^2-4q^2 z)^2}\,.
\eeq

If $q=0$, we could isolate the metric function as $f(z)=\frac{  R_{abcd;e}  R^{abcd;e}}{8  (R_{abcd}R^{abcd} - R_{ab}R^{ab})}$.

 Instead, if we set $M=0$, then we get $f(z)=\frac{  R_{abcd;e}  R^{abcd;e}}{10  (R_{abcd}R^{abcd} - R_{ab}R^{ab})}=\frac{2 (R_{abcd}R^{abcd} - R_{ab}R^{ab}) }{G_{ab;c} R^{ab;c}} $.

 Also, the Ricci scalar is found as
\beq
R=-2(qz)^2+2M^2 z-6\Lambda\,.
\eeq

Thus, we could derive
\beq
R_{;a}R^{;a}=4 z^2 f(z) (M^2 -2 q^2 z)^2\,.
\eeq

If $q=0$, we get the syzygy\footnote{A syzygy is actually a certain algebraic constraint between the curvature invariants of a chosen spacetime \cite{Stephani:2003tm}.}
\beq
R_{abcd;e} R^{abcd;e}-R_{;a} R^{;a}=0\,,
\eeq
while if we only set $M=0$ we would get the following syzygy 
\beq
\label{mass1}
R_{abcd;e} R^{abcd;e}- 5 R_{;a} R^{;a}=0\,.
\eeq

This syzygy would then allow us to define a new parameter which  in section (\ref{sub}), we dubbed {\it "massiveness"}.

 The curvature syzygies actually depend on the matter content of the spacetime, which is in agreement with the field equations delivering different curvatures for different types of fluids. It also depends on the chosen theory of gravity. 

Without any restriction on the value of the electric charge we would then get the following relation
\beq
R_{abcd;e} R^{abcd;e}-4 G_{ab;c} G^{ab;c}=0\,.
\eeq

Note that using the relation $R=g^{ab} R_{ab}$, and the Leibniz rule for the covariant derivative, and also the fact that the covariant derivative of the metric tensor is zero, we could rewrite the previous constraints in other equivalent forms.

Moreover, a second syzygy would be found as
\beq
2(R_{ab} G^{ab}+R_{ab} R^{ab})- R_{abcd} R^{abcd}=0\,.
\eeq

Interestingly, we can get the relation ${\bf e}_z(R_{tt})=\frac{M^2 z}{4} \sqrt{2f(z)}$, which points to the fact that in the massive gravity theory, this specific Cartan curvature invariant could also be used for locating the horizon of the BTZ black hole.

The previous syzygies could also be complemented by the following first order and zeroth order constraints 
\begin{eqnarray}
&& {\bf e}_x (R_{xzzt}) - {\bf e}_x (R_{xz})\,=\, 0, \\
&& a R_{tztz} + b R_{xztx} +c R_{tz} + d R_{xx} +e R\,=\,0, \\
&& a=-2(b+2d+e)\,, \qquad c=2(b +2d+2e),
\end{eqnarray}
which involve Cartan curvature invariants, rather than scalar polynomial curvature invariants.

Moreover, we could write the following relation 
\beq
\frac{{\bf e}_t (R_{tz})}{{\mathcal J}_1} \equiv \frac{M^2}{4z q^2},
\eeq
which actually invariantly quantifies the ratio of the strength of the graviton field over the electromagnetic field. 

\begin{figure}
	\begin{center}
		$
		\begin{array}{cc}
		{\includegraphics[scale=0.4, angle=0]{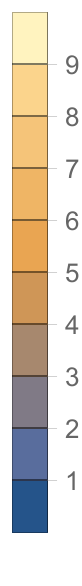}}
		\hspace{0.5mm}
		{\includegraphics[scale=0.4, angle=0]{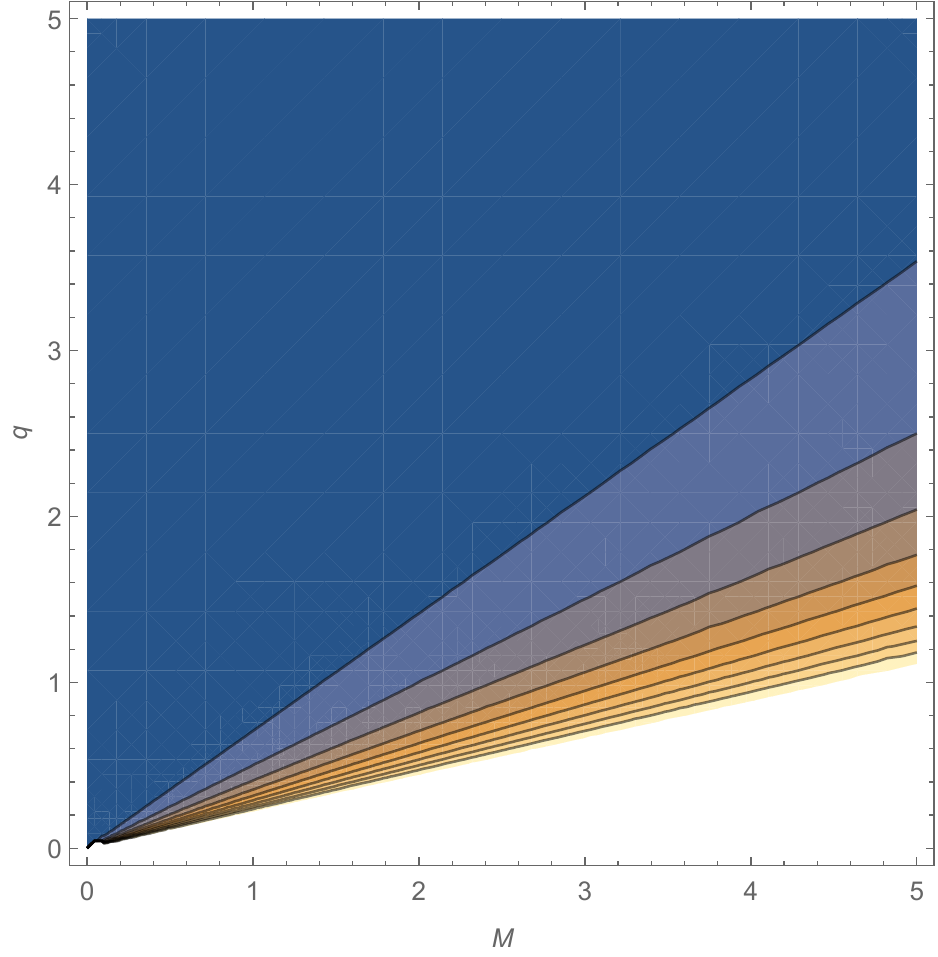}}
		\hspace{3mm}
		{\includegraphics[scale=0.4, angle=0] {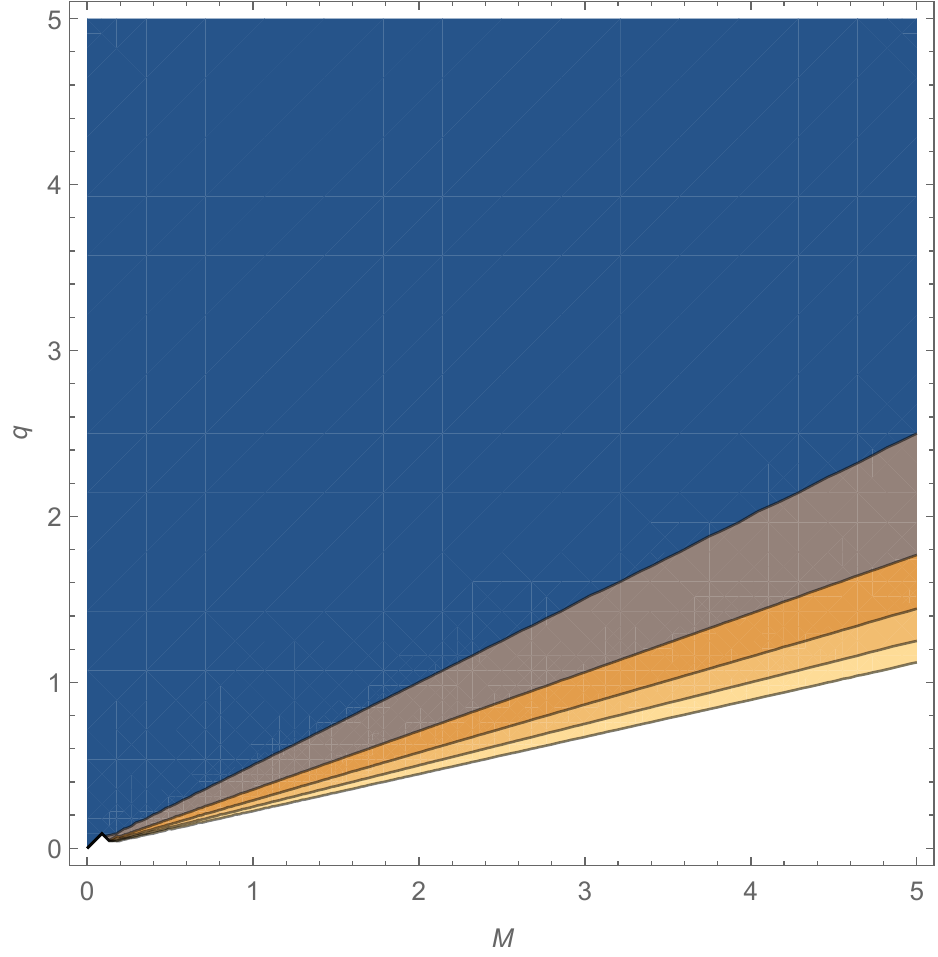}}  \hspace{0.5mm} {\includegraphics[scale=0.4, angle=0]{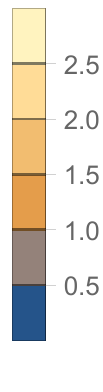}} \\
		(a) \hspace{40mm} (b)  \\[7mm]
		{\includegraphics[scale=0.4, angle=0]{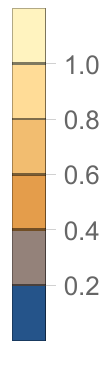}}
		\hspace{0.5mm} {\includegraphics[scale=0.4]{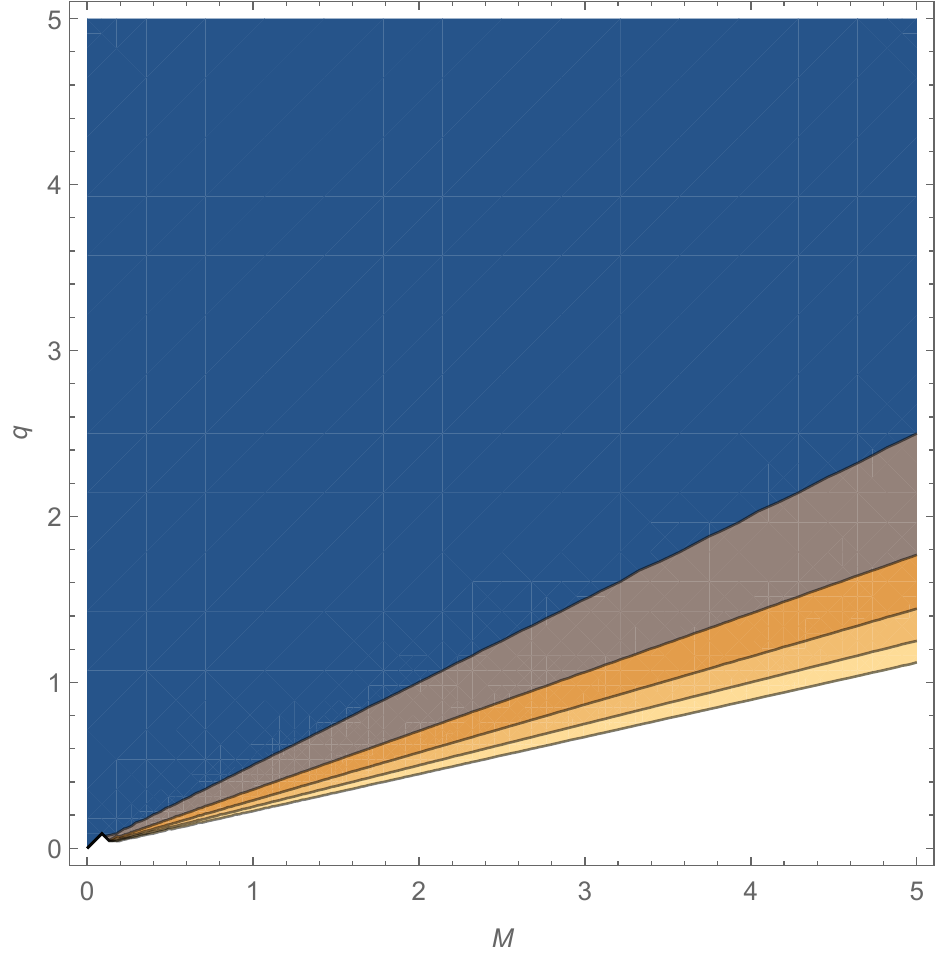}}
		\hspace{3mm}
		{\includegraphics[scale=0.4]{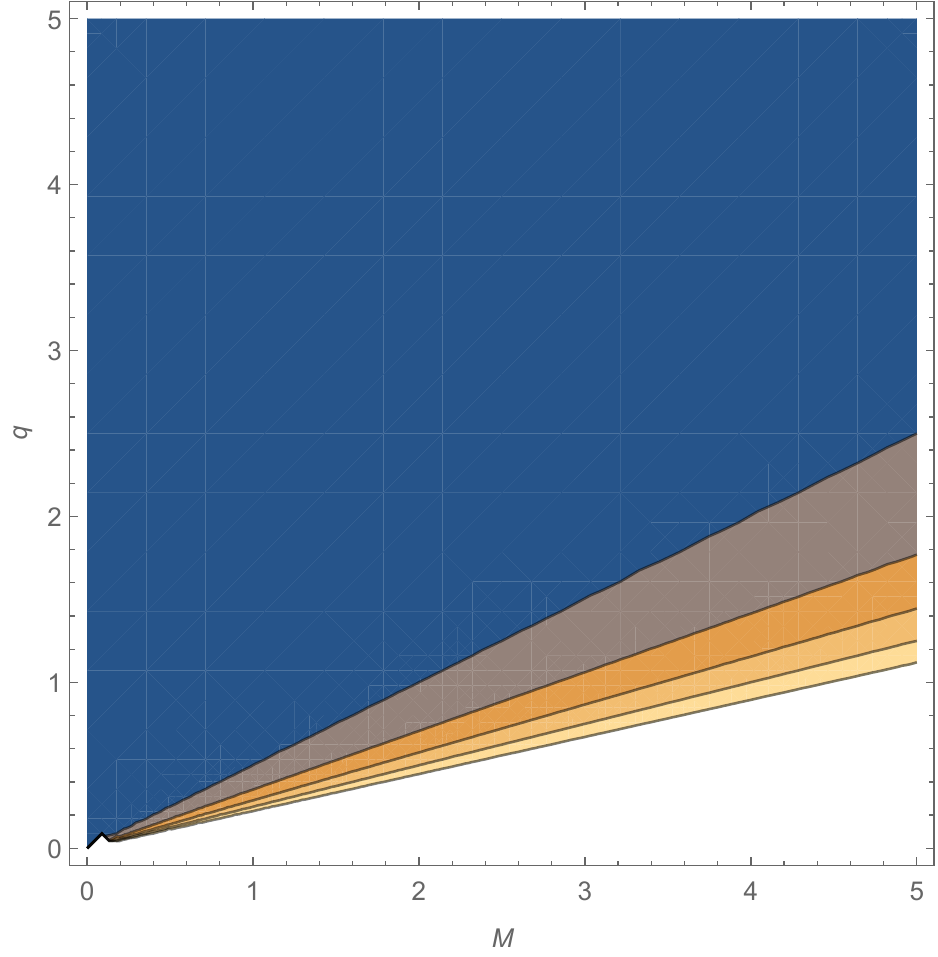}} \hspace{0.5mm} {\includegraphics[scale=0.4, angle=0]{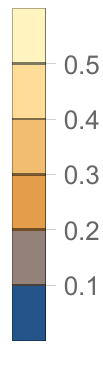}} \\
		(c) \hspace{40mm} (d) 
		\end{array}
		$
	\end{center}
	\caption{In the above figure,  we investigate numerically the relative strength between the two matter field generated by the gravitons and by the Maxwell electromagnetic field. This term is quantified through the Cartan curvature invariant $\frac{{\bf e}_t (R_{tz})}{{\mathcal J}_1}$ at various distances from the center of the black hole as in panel (a) $z=0.5$, in panel (b) $z=2.0$, in panel (c) $z=5.0$, and in panel (d) $z=10.0$. 
	}
	\label{str}
\end{figure}

 In Fig. (\ref{str}), we numerically found the behavior of the ratio of these two matter fields at various distances from the center of the configuration. The ratio of these two fields actually shape the specific properties of spacetime. Note that these plots have been constructed at various distances which are $z=0.5$ in panel (a), $z=2.0$ in panel (b), $z=5.0$ in panel (c), and $z=10.0$ in panel (d).

{Our analysis about the curvature structure of the massive black hole spacetime may help also in clarifying which theory the solution belongs to. Suppose for example that, for some reasons about which we do not make hypotheses, the same metric tensor $g_{\mu\nu}$ is a solution of two systems of differential equations which follow from two different action principles. Let us write such field equations as $G_{\mu\nu}=T_{\mu\nu}$ and  ${\tilde G}_{\mu\nu}=T_{\mu\nu}$, where $G_{\mu\nu}$ and ${\tilde G}_{\mu\nu}$ are some \lq\lq generalized" Einstein tensors which are different functions of the various curvature tensors, and $T_{\mu\nu}$ is the stress-energy tensor. Starting from these tensorial equations, we can construct a number of invariant conditions like $G_{\mu\nu}G^{\mu\nu} =T_{\mu\nu}T^{\mu\nu}$, $G^{\sigma}{}_{\mu}G_{\sigma\nu}G^{\mu\nu} =T^{\sigma}{}_{\mu}T_{\sigma\nu}T^{\mu\nu}$, etc..., and similarly for the tilded system of equations. Then, implementing our results about the dependence of the curvature invariants on the black hole parameters on the left hand side, and the matter sources, like energy denisty and pressure, on the right hand side, we can derive some relationships between the former and the latter which depend on the field equations, that is, on the gravitational theory on which they have been derived. }

\section{Isolating the black hole parameters} \label{s3}

It is a known fact that, for evaluating the parameters of a black hole, the standard procedure in numerical relativity would require extracting them from the area of the horizon \cite{Baumgarte:2010ndz}.

This procedure, conceptually is unsatisfactory, since it would lead to a non-local parameter, while the field theory of gravitation is local by construction. The other problem with this method is that it is computationally very expensive because the computing of the area relies on evaluating an integral. 

With these problems in mind, through some scalar polynomial curvature invariants, the authors of \cite{Abdelqader:2014vaa} proposed a local procedure for writing the mass and angular momentum of a Kerr black hole.

In this section, we will show that it would be possible to isolate the parameters characterizing the black hole solution (\ref{metric}), as some appropriate algebraic expressions involving both scalar polynomial curvature invariants and the specific Cartan invariant (\ref{phi11}).

So first, we begin by defining the ``Kretschmann scalar" as
\beq
K:=R_{abcd} R^{abcd}=2[6(qz)^4-4(Mqz)^2 z+4\Lambda(qz)^2 +(M^2 z)^2 -4\Lambda M^2 z+ 6\Lambda^2]\,.
\eeq

As we expected, for large values of radial coordinate, i.e.  $z\to \infty$, where the black hole is actually located, the Kretschmann scalar would diverge. Also, asymptotically, for the case of $z=0$, it would reduce to the case of a de-Sitter universe. 

These parameters would also be affected by the value of the mass of graviton. However, for the same reason discussed about the effects of various parameters on $\Phi_{11}$, the mass of the black hole cannot change those scalar quantities. Again, this latter effect would be different in the cases of higher-dimensional gravity theories. This  follows directly from the fact that there is no Weyl tensor in the (2+1)-gravity.

Now we define the following auxiliary curvature invariant quantities,
\begin{eqnarray}
{\mathcal I}_2 &=& \left(\frac12(3K -R^2)  \right)^{1/2}, \\
{\mathcal I}_3 &=&\left(- \frac{{\mathcal I}_2 \cdot R_{;a} R^{;a}}{R_{ab;c} G^{ab;c}}  \right)^{1/2}, \\
{\mathcal I}_4 &=& \frac{{\mathcal I}_2 }{4 R_{ab;c} G^{ab;c}}({\mathcal I}_1 - 5  R_{;a} R^{;a}), \\
\label{I5}
{\mathcal I}_5 &=& \frac{3 {\mathcal I}_3- (9 {\mathcal I}_3^2 - 8{\mathcal I}_4)^{1/2} }{2} \equiv \frac{M^2}{q}, \\
\label{I6}
{\mathcal I}_6 &=& \frac{{\mathcal I}_5 ^2 - {\mathcal I}_4 }{3}   \equiv M^2 z, \\
{\mathcal I}_7 &=& \frac{1}{4 {\mathcal I}_2 ^2} \left(K \cdot R -\frac{R_{ac}R^{a}{}_{b}R^{bc}}{3}  \right) \equiv M^2 z - 4 \Lambda\,.
\end{eqnarray}

In addition, the cosmological constant for this massive BTZ solution could be written locally and invariantly as
\beq
\Lambda=\frac{{\mathcal I}_6 - {\mathcal I}_7}{4}\,.
\eeq

In the next step, we define the following quantity
\beq
{\mathcal I}_8 = \frac{R_{ab;c} G^{ab;c}}{4 \left(\frac{{\mathcal I}_6}{{\mathcal I}_5}\right)^2\left[{\mathcal I}_6 -4 \left(\frac{{\mathcal I}_6}{{\mathcal I}_5}\right)^2 \right]} \equiv f(z).
\eeq

Using the relation below written for the massive BTZ solution,
\beq
-\left(  {\mathcal I}_8-\frac{{\mathcal I}_7 + 3{\mathcal I}_6}{4} \right)\equiv mz^2 -2 q^2 z^2 \ln (z)=m \left( \frac{{\mathcal I}_6}{q {\mathcal I}_5} \right)^2 - 2\left(\frac{{\mathcal I}_6}{ {\mathcal I}_5}\right)^2 \ln\left( \frac{{\mathcal I}_6}{q {\mathcal I}_5}\right)\,,
\eeq
we could get the mass of  the black hole in terms of the curvature invariants as
\beq
m=\left[-\left(  {\mathcal I}_8-\frac{{\mathcal I}_7 + 3{\mathcal I}_6}{4} \right) + 2\left(\frac{{\mathcal I}_6}{ {\mathcal I}_5}\right)^2 \ln\left( \frac{{\mathcal I}_6}{ q{\mathcal I}_5}\right)\right]\left(\frac{q{\mathcal I}_5}{ {\mathcal I}_6}\right)^2\,.
\eeq

Note that a logarithmic factor appears here as one would expect for such solutions. Thus, as we could relate the mass of the  the black hole as this specific combination of the curvature quantities, our loop is closed and therefore the massive BTZ black hole would be fully characterized ``locally".

 As for finding the mass of the graviton field in terms of the curvature invariants we would have
\begin{eqnarray}
3 \Phi_{11}+{\mathcal I}_7 &\,=\,&-\frac{q^2}{2z^2}+\frac{M^2}{4 z^3}-\frac{{\mathcal I}_6 - {\mathcal I}_7}{8 z^2}\\
&=& \left(\frac{1}{4 {\mathcal I}_6^3}-\frac{1}{2{\mathcal I}_5^2{\mathcal I}_6^2}  \right)M^8 - \frac{{\mathcal I}_6 - {\mathcal I}_7}{8 }\frac{M^4}{{\mathcal I}_6^2}\,,
\end{eqnarray}
where in the second step we used the relation (\ref{I6}). In non-massive gravity theory, the right hand side of this equation would actually vanish.

Considering this point, we could propose a new quantity which we dub the {\it massiveness} parameter.  We discuss about the properties of this parameter further in section (\ref{sub}).

 For the mass  of the graviton, introducing the auxiliary variable $\mu:=M^4$, we would get the second degree equation as
\beq \label{eq:curvatureEQ}
2({\mathcal I}_5^2 -2 {\mathcal I}_6)\mu^2 -({\mathcal I}_6 - {\mathcal I}_7)\,{\mathcal I}_6^3 \,{\mathcal I}_5^2 \,\mu - 8\,{\mathcal I}_6^3\,{\mathcal I}_5^2\,(3 \Phi_{11}+{\mathcal I}_7)=0\,.
\eeq

Therefore, the mass of  the graviton field could also be written in terms of our specific curvature invariants as we wanted.
 
 We will later comment on the physical meaning of these terms in further details.

\subsection{The massiveness parameter}
\label{sub}

Our investigations of the curvature properties and also the specific parameters of the massive BTZ black hole could lead us to introduce a {\it massiveness} parameter ${\mathcal M}$ which measures the {\it distance} between a lower-dimensional black hole solution in the massive gravity and the one in regular gravity in terms of the scale of energy.

This parameter could act as a geometrical measure for specifying the  \textit {massiveness} of the graviton and its effects on the background solution, or how much different the theory would be from the Einstein gravity. 

Using this quantity, a similar heat map contour of spacetime could also be constructed, which could be used as  another tool to visualize these massive gravity models better. This quantity could also show better the physical regions of the theory where the calculations could completely be trusted.

In fact, this procedure followed the same line of thinking pioneered in \cite{Abdelqader:2014vaa} which has introduced a {\it kerness} parameter which for astrophysical black holes, would actually measure the distortions and the deviations from the axis-symmetry.

We could propose two definitions for this ``massiveness" quantity. One would involve only the scalar polynomial curvature invariants, and the second one would involve a mixture of Cartan and scalar polynomial curvature invariants. These two definitions would be as follows
\begin{eqnarray}
{\mathcal M}_1&:=&R_{abcd;e} R^{abcd;e}- 5 R_{;a} R^{;a}, \\
{\mathcal M}_2&:=& 3 \Phi_{11}+{\mathcal I}_7\,.
\end{eqnarray}

\begin{figure}
\begin{center}
    $
    \begin{array}{cc}
    {\includegraphics[scale=0.4, angle=0]{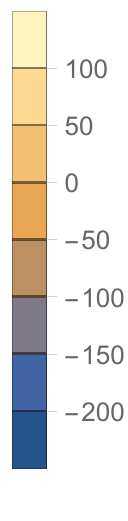}}
\hspace{0.5mm}
{\includegraphics[scale=0.4, angle=0]{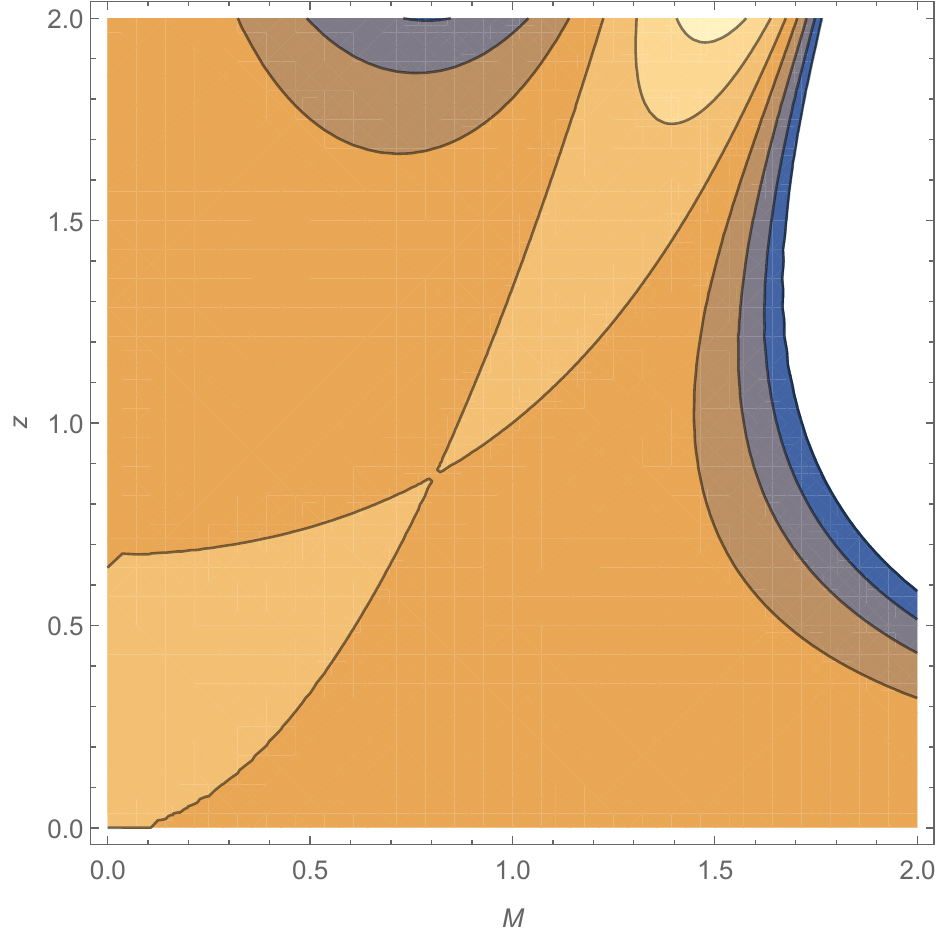}}
\hspace{3mm}
{\includegraphics[scale=0.4, angle=0] {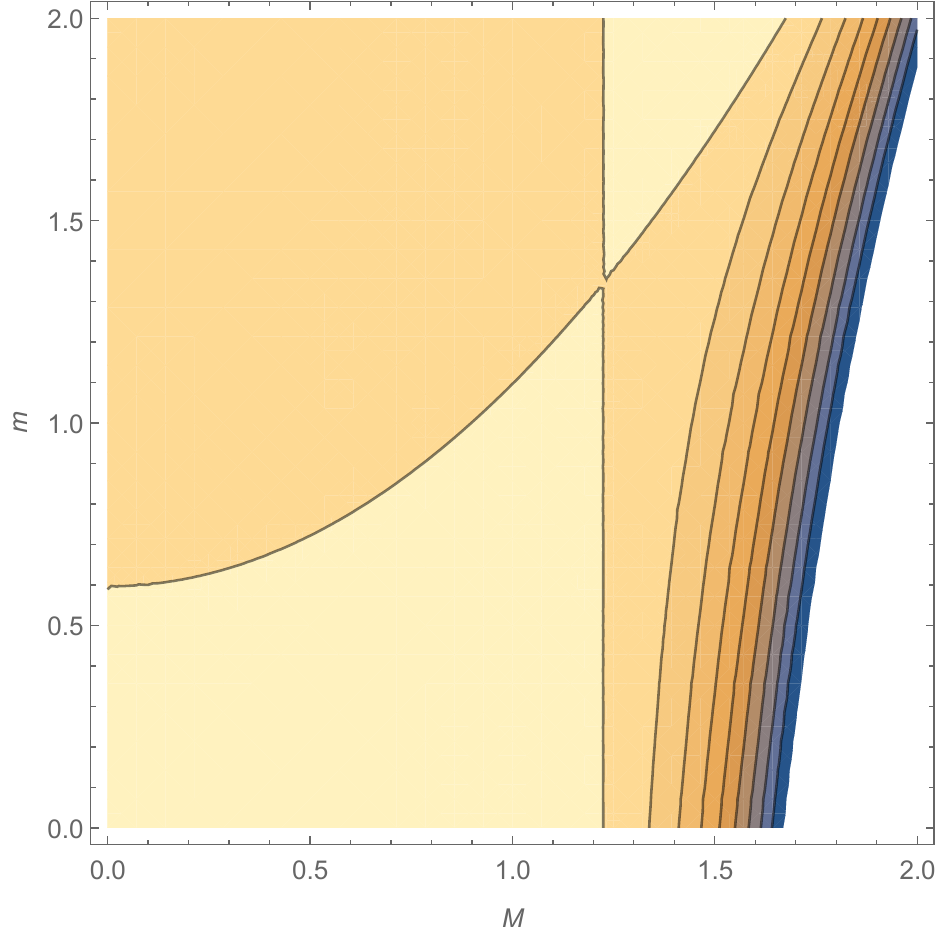}}  \hspace{0.5mm} {\includegraphics[scale=0.4, angle=0]{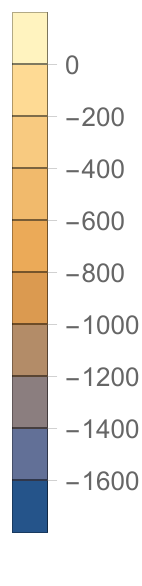}} \\
(a) \hspace{40mm} (b)  \\[7mm]
{\includegraphics[scale=0.4, angle=0]{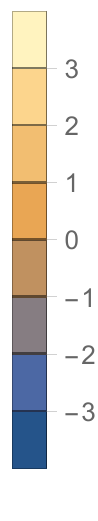}}
\hspace{0.5mm} {\includegraphics[scale=0.4]{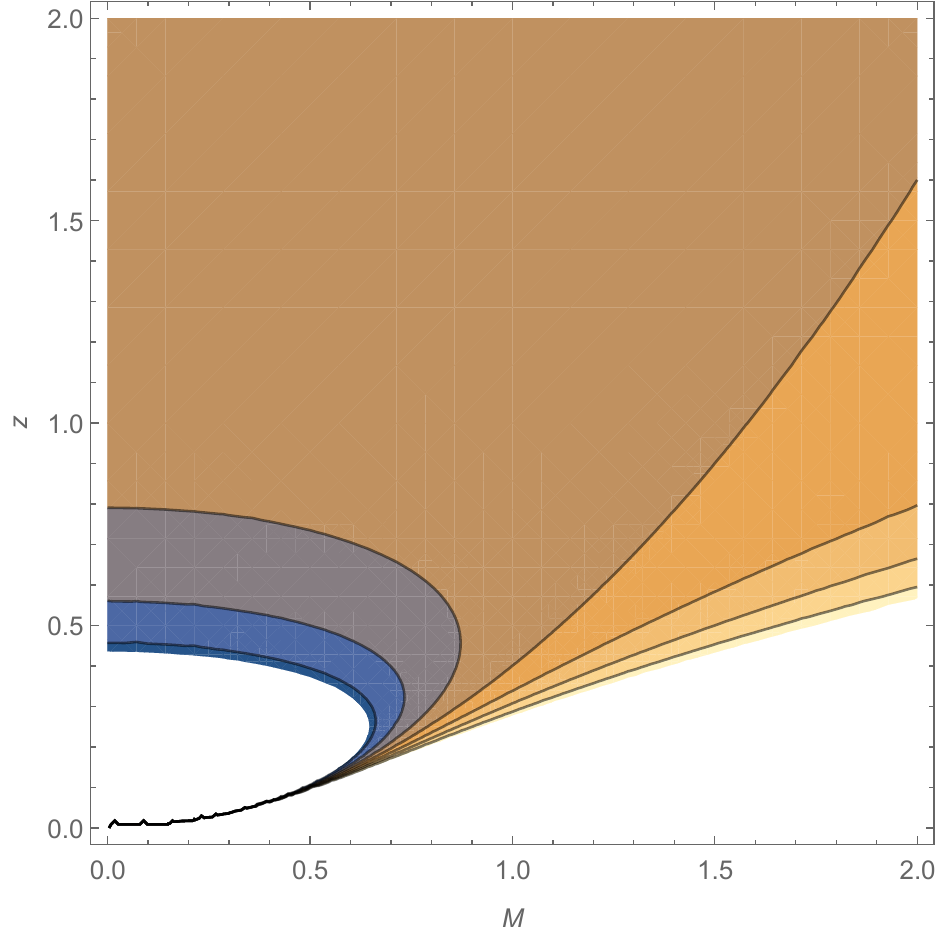}}
\hspace{3mm}
{\includegraphics[scale=0.4]{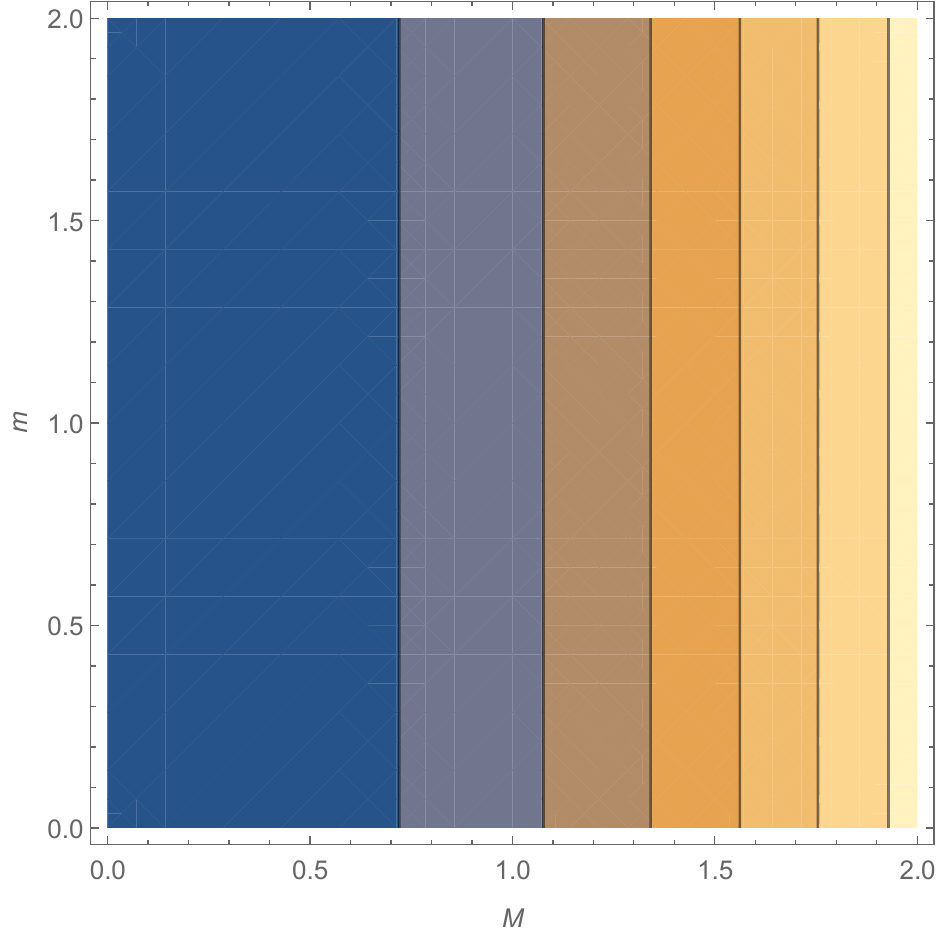}} \hspace{0.5mm} {\includegraphics[scale=0.4, angle=0]{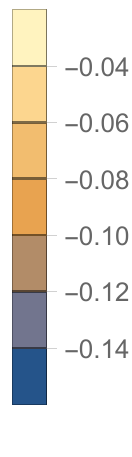}} \\
 (c) \hspace{40mm} (d) 
\end{array}
$
\end{center}
\caption{The figure  shows the contourplots of the two massiveness parameters ${\mathcal M}_1$ in panels (a)-(b), and ${\mathcal M}_2$ in panels (c)-(d) in the parameter spaces ($z$, $M$) and ($m$, $M$) respectively. We have chosen the remaining free parameters to be ($\Lambda=-1$, $q=0.5$, $z=2.0$) and ($\Lambda=-1$, $q=0.5$, $z=2.0$). 
}
\label{mass}
\end{figure}

In Fig. (\ref{mass}), we compare and contrast the behaviors of these two quantities as contour-plots.  In panels (a)-(c), the parameter space would be ($z$ and $M$), and in panels (b)-(d), the parameters are ($m$ and $M$).

\section{The connections between global and local techniques for locating black hole horizon}
\label{s4}

The most widely used technique for finding the locations of the black hole horizons relies on the analysis of the focusing properties of a bundle of light rays. In fact,  the expansion of a bundle of light rays, ``on the horizons" would be zero and as it could also be seen by studying the Raychaudhuri differential equation \cite{Misner:1974qy,Ashtekar:2004cn,Bousso:2002ju}, it switches its sign when crossing that hypersurface. 

The Raychaudhuri equation would actually account for the evolution of the expansion of a bundle of light rays in terms of several parameters which are the expansion itself, the shear and vorticity of the bundle, and the projection of the Ricci tensor in the direction of the null geodesic vector.

This approach could identify the black hole horizon with a marginally outer trapped surfaces (MOTS), see for example chapter 7 in \cite{Baumgarte:2010ndz}. Note that, while for static and stationary spacetimes, MOTS and black hole horizons constitute exactly the same hypersurface within the manifold (like in the case of this paper) \cite{Gibbons:1972ym}, in the case of dynamically evolving configurations, MOTS could only be considered as an approximation of the black hole horizon \cite{Ashtekar:2004cn}.

Recently, in a number of papers, the evolution of multi-black-hole systems in the mathematical and numerical relativity setups have been studied, and adopting the concept of MOTS, the locations of the black hole horizons have been approximated, see for instance \cite{Bentivegna:2012ei,Clifton:2013jpa,Yoo:2013yea}. Moreover, it could be shown that MOTS would be the minimal surfaces within the manifold (i.e. surfaces which minimize the area functional) \cite{Gibbons:1972ym}.

However, one should note that both of these techniques, which search for the horizon in terms of a MOTS or as a minimal surface, are computationally very expensive. This is due to the fact that they rely on the solution of a differential equation (for the former case) or the computation of an integral of the area functional (for the latter case). Moreover, they would need information on more than one spacetime point. Therefore, our local methods presented here, which rely only on the algebraic equations, would be much more preferable.

In this section, we establish the equivalence between the Raychaudhuri and Cartan methods. We could do this by connecting the latter curvature invariant we have constructed for locating the horizon, to the causal properties of the full massive BTZ black hole spacetime.

Therefore, first, let us set the 4-vector $k^\alpha= F(t,z)[1,\,f(z),\,0]$ to be a null geodesic vector (i.e. the function $F(t,z)$ must be determined by integrating the geodesic equations of motion)\footnote{While for our purposes here we do not need any sharper information, for completeness it worths to mention that a deeper investigation of the geodesic motion in a BTZ spacetime in massive gravity has been carried out in \cite{Hendi:2020yah}.}. The authors of \cite{Clifton:2013jpa}, in the section 6.1 of their work, could prove that {\it  on a time-symmetric hypersurface, MOTS are extrinsically flat, and hence with indeterminate lines of curvature would be totally geodesic}. 

Since a direct computation for the spacetime (\ref{metric}) would deliver the relation, $R_{\alpha\beta}k^\alpha k^\beta=0$, and recalling the fact that staticity property of solution would imply the time-symmetry at any moment, we could then use the same hypotheses of \cite{Clifton:2013jpa} as all the conditions would be satisfied here. Thus, the horizon of a static and spherically symmetric massive BTZ black hole solution would be a hypersurface with indeterminate lines of curvature. Moreover, the existence of the spherical symmetry in the spacetime (\ref{metric}) would imply a diagonal Ricci tensor. Therefore, the vector ${\bf e}_z$ would be a principal direction of the Ricci tensor as we have the relation $R_{tz}=R_{zx}=0$.

 As the result of the above arguments, we could obtain the following relations, $0={\bf e}_z(R^{zz})={\bf e}_z(R^{tzz}{}_{t}+R^{zzz}{}_{z}+R^{xzz}{}_{x})={\bf e}_z(R^{tzz}{}_{t})+{\bf e}_z(R^{zzz}{}_{z})+{\bf e}_z(R^{xzz}{}_{x})={\mathcal J}_1$, which connect the local and global techniques of finding horizon for the case of massive BTZ black hole.

\section{The irreducible mass of the massive BTZ black hole }
\label{s5}

Solving the relation appearing among equations (3.15)-(3.16) in \cite{Hendi:2016pvx}, which is in the form of $m=m(M,q,\Lambda)$, would deliver the irreducible mass of a massive BTZ black hole. { Since we are dealing with an isolated and stationary black hole, its mass parameter and its ADM mass correspond  \cite{papantonopoulos:2009bh}.}
 After that, other thermodynamical properties of black holes could be obtained.

In reference \cite{Hendi:2016pvx}, the authors applied the Hawking proposal, which conjectures that the area (or equivalently the entropy) of the black hole horizon would have a non-decreasing behavior by passing time \cite{Hawking:1971tu}. The same result could be obtained by applying the concept of reversible versus irreversible black holes transformations proposed by Christodoulou and Ruffini \cite{Christodoulou:1970wf,Christodoulou:1972kt}.  

We actually noted that the reference \cite{Hendi:2016pvx} did not cite any of these two papers, and thus it seems to us that in their work, they actually did not recognize the full physical interpretations and consequences of the Christodoulou and Ruffini's work. 

The Christodoulou-Ruffini proposal relies on the study of the motion of a test-particle of mass $\mu$, energy $p_t=E$, radial momentum $p_z$, and angular momentum $p_x$ whose normalization condition $g_{ab}p^a p^b=-\mu^2$ reads as
\beq
(z E)^2 - (z f(z) p_z)^2  - (z p_x)^2 f(z) -f(z) \mu^2=0\,.
\eeq

The notion of the Christodoulou-Ruffini irreducible mass has  been proposed originally for rotating Kerr black hole \cite{Christodoulou:1970wf} and then extended to the charged Kerr-Newman solution \cite{Christodoulou:1972kt}. The key physical step was the Penrose process of particle-antiparticle creation \cite{Penrose:1969pc}, with the subsequent  absorption of the antiparticle and energy extraction from the black hole. Although the Penrose process occurs in the proximity of the black hole ergosphere, which exists solely for rotating spacetimes, the irreducible mass $M_{\rm irr}$ differs from the black hole mass parameter $m$ entering the metric tensor even for non-rotating black holes: this can be appreciated  from the  zero angular momentum limit applied to the result in \cite{Christodoulou:1972kt}. Indeed,  the irreducible Christodoulou-Ruffini mass has been computed also for non-rotating black holes \cite{Pereira:2014xua,Pereira:2015pva}, as it may serve for quantifying the energy which can be extracted from the black hole by emission of uncharged scalar particles. Hence, we will now discuss the irreducible mass for the non-rotating spacetime  (\ref{metric}).

\begin{figure} 
	\begin{center}
		$
		\begin{array}{cc}
		{\includegraphics[scale=0.30, angle=0]{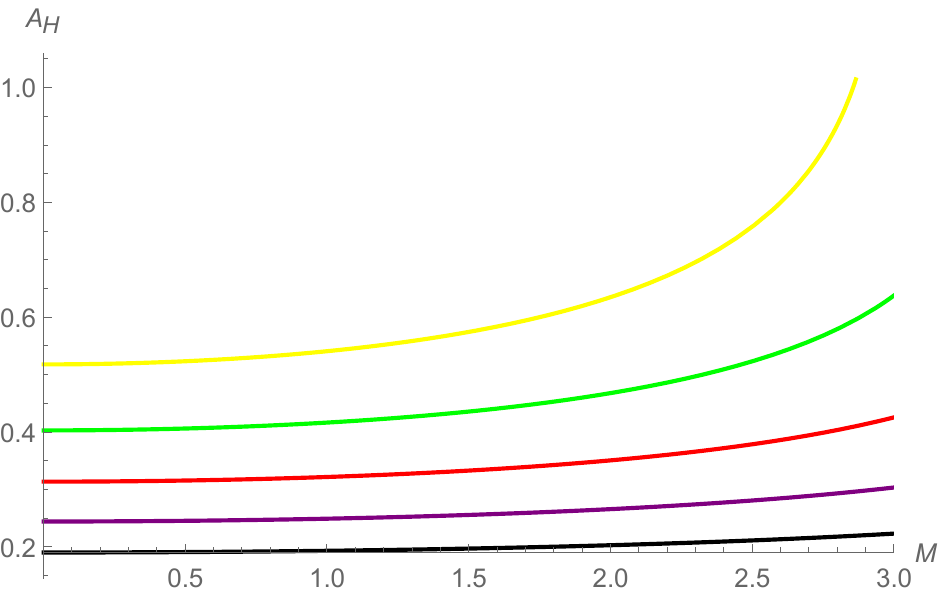}}
		\hspace{2.2 mm}
		{\includegraphics[scale=0.30, angle=0]{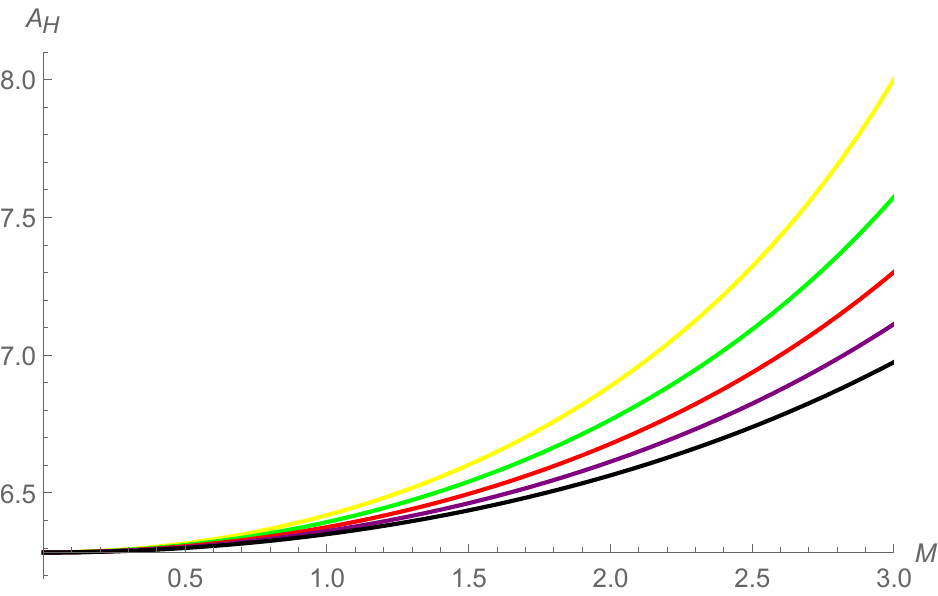}}
		\hspace{2.2 mm}
		{\includegraphics[scale=0.30]{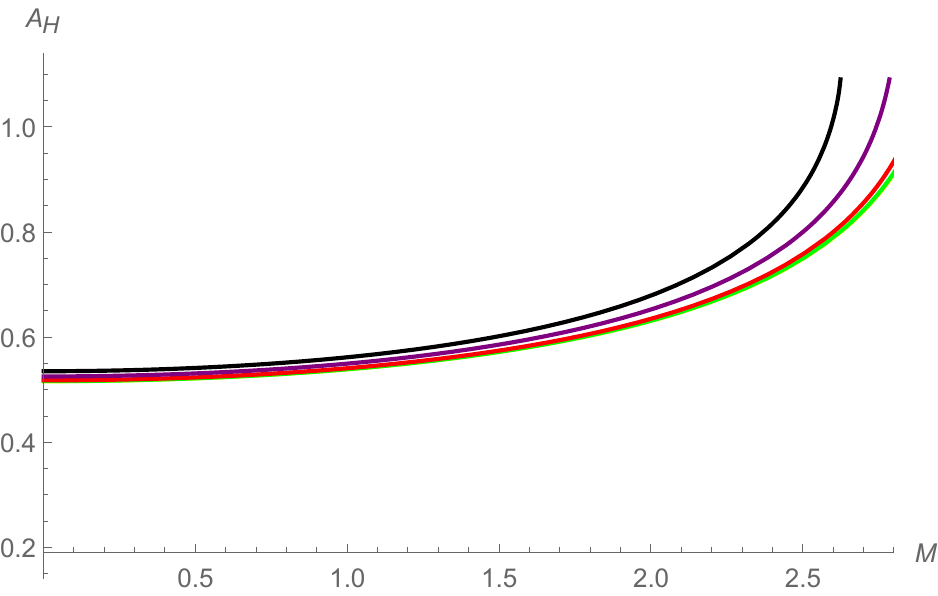}}\\
		(a) \hspace{45mm} (b) \hspace{45mm} (c)  
		\end{array}
		$
	\end{center}
	\caption{The figure depicts the effects of the mass of the graviton field on the area $A_H$ of the outer horizon. In panel (a) we have fixed $\Lambda=-1.0$, $q=1.0$, and from the bottom to the top, the values of mass are $m=5.0$ (black line), $m=5.5$ (purple line), $m=6.0$ (red line), $m=6.5$ (green line), $m=7.0$ (yellow line), showing that the value of the area decreases for more massive black holes. In panel (b) we have fixed $\Lambda=-1.0$, $m=1.0$, and from the bottom to the top the charges are $q=5.0$ (black line), $q=5.5$ (purple line), $q=6.0$ (red line), $q=6.5$ (green line), $q=7.0$ (yellow line), showing that the electric charge has a similar effect than the mass of the areal size of the black hole. Finally in panel (c) we have fixed $m=5.0$, $q=1.0$, and from the bottom to the top, the cosmological constants are $\Lambda=-0.01$ (yellow line), $\Lambda=-0.10$ (green line), $\Lambda=-1.00$ (red line), $\Lambda=-5.00$ (purple line), $\Lambda=-10.00$ (black line), showing that the value of the area increases when the strength of the cosmological constant increases. Moreover, the mass of the graviton field acts similarly to the cosmological constant, but mitigating the roles of the mass and of the electric charge of the black hole. }
	\label{figarearea}
\end{figure}

Therefore, it is important to further depict the physical properties of the horizon in proximity of which particles' emission may occur, as we do in Fig. (\ref{figarearea}). In the reference \cite{Hendi:2016pvx}, the outer horizon, has been identified as the relevant surface on which the Hawking temperature should be estimated. This then implys that this is actually the location where the absorption process takes place.

Note that the area of such horizon is $\mathcal{A}_{ H}=\frac{2 \pi}{z_+}$  where $z_+$  is the larger root of the equation $f(z)=0$. Thus, keeping in mind the definition of the radial coordinate $z$, as intuitively expected, we confirm that the area is increasing when the cosmological parameters (mass of the gravitons, and cosmological constant) are increased.   This quantity increases as well for more massive black holes and also when the Maxwell electromagnetic field becomes stronger.

These results have been shown in more details in Fig. (\ref{figarearea}). In the panel (a) we have fixed $\Lambda=-1.0$, $q=1.0$, and from the bottom to top the curves correspond to different values of mass as $m=5.0$ (black line), $m=5.5$ (purple line), $m=6.0$ (red line), $m=6.5$ (green line), $m=7.0$ (yellow line).

 In panel (b) we have fixed $\Lambda=-1.0$, $m=1.0$, and from the bottom to top the values of charge are, $q=5.0$ (black line), $q=5.5$ (purple line), $q=6.0$ (red line), $q=6.5$ (green line), $q=7.0$ (yellow line).
 
 In panel (c) we have fixed $m=5.0$, $q=1.0$, and from the bottom to top, the values for the cosmological constant are, $\Lambda=-0.01$ (yellow line), $\Lambda=-0.10$ (green line), $\Lambda=-1.00$ (red line), $\Lambda=-5.00$ (purple line), $\Lambda=-10.00$ (black line).
 
 The above results would help to find the cases with a larger surfaces which are also more favorable. These then would be available for the particles' emission phenomenon.

\section{The holographic picture of curvature invariants and the applications}\label{sec:holographyDisc}

One could also consider the dual pictures of the defined curvature invariants and the syzygies on the boundary CFTs, and the applications they might have in the field of AdS/QCD, quantum information and also resolving the information paradox of black holes.

From the holographic point of view, in fact, for any phenomena and characteristics of black holes, one could imagine a dual property in the boundary CFT side. So first, consider that, one of these phenomenon occurring in black holes, is actually the energy extractions. For the non-rotating black holes, the mass-energy formula for a black hole is clarified in \cite{Ruffini:2002rk} and then later it was extended to rotating cases \cite{Kaya:2007zz} and to Einstein-Born-Infeld black hole in \cite{Breton:2016qyf}. The energy that could be extracted from the black hole is holographically dual to the energy one could extract from the confined QCD phase of matter, and the plasma expansion in the geometry of a collapsing star \cite{Ruffini:2003yu} is in fact dual to the behavior of non-confined phase around the central confined part of the quark-gluon plasma of the boundary CFT. In fact, for various metrics such as our case of massive black hole, these geometries could be derived and the effects of various quantities such as mass of graviton which could quantify the momentum dissipations in such geometries, could be studied. 

Note also that the irreducible mass $M_{\text{irr}}$ is actually the energy that cannot be extracted from the black hole via classic processes. This irreducible part cannot get lowered by the classical (non-quantum) processes while the rest such as rotational or Coulomb energy could be extracted by the physical process such as electro-dynamical processes or superradiance (which is similar to stimulated emission in atomic physics). This two parts then could be related to the classical versus quantum correlations in the dual boundary CFT sides. The quantum correlations such as entanglement and quantum discord could then holographically be connected to the irreducible mass $M_{\text{irr}}$.  In addition, in \cite{Ghodrati:2019hnn}, the complexity of purification (CoP) versus interval volume (VI) have been introduced where the former characterizes the complexity growth rate mainly due to classical correlations while the later could give a quantification of the ``quantum'' correlations among the mixed states. Therefore, we conjecture the parameters such as entanglement and quantum discord are holographically related to the irreducible mass of the black hole.

To consider this conjecture in more details, note that as Bekenstein derived in \cite{Bekenstein:1971ej} and also from equation 8 of \cite{Ruffini:2002rk},
\begin{gather}
M_{\rm irr}=m_0-\frac{m_0^2}{2r_+}+T_+, \ \ \ \ \ \ \text{while} \  \ \ m=M_{\rm irr}+\frac{Q^2}{2r_+},
\end{gather}
the irreducible mass $M_{\rm irr}$ is independent of the electromagnetic energy and is just a combination of the rest mass, gravitational potential energy and the kinetic energy of the rest mass evaluated at the horizon. Non of these could be lowered then by a classic process and only could be reduced by a quantum process.

In the dual boundary side, then each of these terms is dual to a ``quantum'' correlation, such as entanglement, quantum discord, quantum deficit, measurement induced disturbance (MID) measure, localized noneffective unitary (LNU) distance, and other various entropy-based measures. We also conjecture that the irreducible mass would be related to the dual entanglement wedge mutual information (EWMI) introduced in \cite{Umemoto:2019jlz}, which is a novel measure of quantum correlations.

In fact, the correlations such as quantum discord is related to correlation in mixed but separable states. One could think that the rest mass of black hole and its surrounding are two mixed but separable states related to each other through the potential energy and then in the holographic dual picture the dual of these two states are correlated via the quantum discord. The entanglement entropy would then be dual to the rest mass.

 Another interesting idea between our work in the bulk gravity side and quantum information in the CFT side, would actually be related to the testing of quantum gravity in the lab \cite{Landsman:2018jpm, Brown:2019hmk}. We could connect the curvature invariants with some dual entanglement structures and its invariant properties, therefore if we could introduce novel quantum circuits which could probe the invariants in the entanglement structures, these would actually probe the dual curvature invariants of the bulk spacetimes.  For instance, one dual picture could be similar to the invariance of multiboundary entanglement entropy (MME), specially for the case of TQFTs, along the orbits of Galois action \cite{Buican:2019evc}, which has been studied there for the case of Abelian and non-Abelian TQFTs. Our algebraic equations then could correspond to similar ``Galois conjugations''-like equations.  Also, other invariants in low-dimensional holography which have been studied in \cite{Rashkov:2019wvw} such as projective invariants, higher projective invariants, Aharonov invariants and their relations to Faber polynomials, Grunsky coefficients and Toda tau-functions, could play the roles in finding the connections between our curvature invariants and syzygies and the entanglement entropy structures. 
 
Furthermore, we propose that our studies could be related to the new resolution of the information paradox of black holes and Page curves \cite{Penington:2019npb, Liu:2020gnp,Nian:2019buz,Chen:2019uhq}.  For instance in \cite{Moffat:2014aqa}, it was conjectured that the behavior of Karlhede's invariant, which is formed by the contraction of the covariant derivative of the Riemann tensor in the form $\mathcal{I}= R^{\alpha \beta \gamma \delta;\kappa} R_{\alpha \beta \gamma \delta;\kappa}$, might indicate that the event horizon is actually a real physical membrane where a freely falling observer could measure its properties, so it advocates for the existence of a firewall. The other curvature invariants for various black hole solutions such as our massive case could also be studied and with this new lens, for each case, the debate of complementarity versus firewall \cite{Almheiri:2012rt}, Page curves \cite{Nian:2019buz} or the effects on the echoes of black holes \cite{Dey:2020lhq}, could be investigated.

Additionally, one should note that the dual picture of the geometrical curvature in the bulk is the renormalization flow in the dual boundary field theory side. Therefore, one would expect that the curvature invariants could be used to find the renormalization flows and specifically the anomalies such as Weyl, trace or conformal anomalies. For instance in \cite{Bugini:2016nvn}, the holographic Weyl anomaly has been found in theories with higher curvature invariants. There, they found the natural basis for the CFT trace anomaly as the one which trades the Euler density with the pure Ricci ``Q-curvature''. Therefore, in this basis, they found that the bulk volume would descend to the Q-curvature and the bulk Weyl invariant to the Weyl invariants of the type-B Weyl anomaly. In other works such as \cite{Natsuume:2019sfp, Taylor:2016aoi,Li:2018drw}, etc, various curvature invariants have also been used to find the renormalization group flows and anomalies in different holographic theories. Therefore, the effects of momentum dissipations on the renormalization group flows could be studied by the curvature invariants  and the syzygies which we have found in this work.

Another interesting result, recently found in \cite{deBoer:2019uem, Czech:2019vih}, which could also point to a holographic application for our study, is that the operators which saturate modular chaos, which has been dubbed in \cite{deBoer:2019uem} as the \textit{modular scrambling modes} would holographically be mapped to local null shifts of the Ryu-Takayanagi (RT) surfaces $\zeta^\pm$, in the bulk, that preserves the normal frame on it. These operators indeed satisfy the relation
\begin{gather}
\mathcal{L}_{\zeta^{\pm}} g_{\mu \nu} \big |_{\text{RT}} \approx 0.
\end{gather}

Note that the commutators of these operators would probe the bulk Riemann curvature.

Therefore, we could say that the bulk curvature invariants would in fact be related to the operators of modular scrambling modes of the boundary field theory side. 

Note that these modes are actually in the form of
\begin{gather}
G_{\pm}= \pm \frac{e^{-2 \pi \Lambda} }{2 \pi}  \Big ( dx_1 dx_2 \delta K_{ab}^{(0)} (x_1,x_2) \phi_a(x_1,\Lambda) \phi_b(x_2,\Lambda)\nonumber\\
+ \frac{1}{N} \int dx_1 dx_2 dx_3 \delta K_{abc}^{(1)} (x_1,x_2,x_3) \phi_a(x_1, \Lambda) \phi_b (x_2, \Lambda) \phi_c(x_3, \Lambda)+ ... \Big ),
\end{gather}  
and these operators satisfy the following relations
\begin{flalign}
[G_+, G_-] &= J^i (y) \partial_i +\delta x^+ \delta x^- {{R_{+- |}}^\mu}_\alpha x^\alpha \partial_\mu\nonumber\\
&+ \Big ( \frac{1}{2} \epsilon_{-+} \nabla^i \delta x^- \nabla_i \delta x^+-2 J^i (y) a_i (y) \Big) (x^+ \partial_+ - x^- \partial_-)\nonumber\\
& \text{where:} \  J^ i =\frac{1}{2} ( \delta x^- \nabla^i \delta x^+ - \delta x^+ \nabla^i  \delta x^- ).
\end{flalign}

For the cases where these modular operators commute or are zero, we could recognize a curvature invariant equation such as equation (\ref{eq:curvatureEQ}) in the bulk side.  As mentioned in \cite{deBoer:2019uem}, these commutators however are non-vanishing when there is the following relation between the extrinsic curvature of the RT surface and bulk curvature as
\begin{gather}
R \gg K^2.
\end{gather}

So indeed there would be a relation between bulk curvature, its invariants and syzygies, the values of $[G_+,G_-]$ and its algebra, and also entanglement structures of the boundary CFT side.

As we consider a massive gravity background,  we are actually considering a field theory with dissipations where they affect the scrambling modes $G_{\pm}$, the boundary algebra and as the result the curvature properties of the bulk and the curvature invariants. As we expect the mass term suppresses the curvature invariants, scrambling modes and syzygies.

Moreover, as quantities such as mass and angular momentum could be extracted from the curvature invariants \cite{Abdelqader:2014vaa}, we expect that in general all the conserved charges of any metric could be written in terms of the specific curvature invariants for that metric. As the symmetries of these conserved charges generate certain algebra, then one could depict the connections between curvature invariants and these algebras on the boundaries. Therefore, the connections between the specific curvature invariants for each metric such as Kerr metric and the dual boundary algebra, for instance the $\text{BMS}_3$-type or the recent T-Witt algebra \cite{Adami:20200j} for the integrable part of surface charges of Kerr metric, or other metrics such as our massive black hole and the corresponding algebra for its conserved charges would be a very interesting line of research for the future studies. 
In addition, the connection between locating ergosurfaces using curvature invariants \cite{Abdelqader:2014vaa}, and non-integrable part of symmetry algebra which accounts for the fluxes which pass through the null hypersurfaces \cite{Adami:20200j} could be studied.  

For other holographic pictures for our studies, we could note that the dual region of ergosurface and ergospheres have been studied in the holographic pictures of rotating black holes \cite{Siahaan:2014ihe,Chen:2019zob}. The same studies then could be done for the rotating massive black holes and then quantities such as OTOC could be calculated for this case.

Besides, the AdS/CFT even has been used in \cite{Hashimoto:2018okj} to construct directly the holographic images of the black hole, and even the Einstein ring \cite{Hashimoto:2019jmw}, from the response functions of the boundary QFTs. The specific curvature invariants have been implied for those studies. Using only curvature invariants and syzygies, one could also build such images. These studies then could be repeated for our massive case and the effect of graviton mass on the shape and size of ergospheres and images of black hole could be further studied. Note, that in gathering the information to obtain the image of black hole by the Event Horizon Telescope (EHT), the environment of the black hole has been scanned as well. This therefore, again points to the necessity of a correct excision method for various solutions, specifically those massive gravity metrics, which have been discussed in more details here.

 Recently in \cite{Tavlayan:2020chf}, a similar method to \cite{Abdelqader:2014vaa} has been proposed which instead of using the full spacetime, they use the curvature invariants of codimension one hypersurface, so by reducing the local cohomogeneity, they could apply their method even for the BTZ case.

 This method could also be used for our massive BTZ black hole and then compare the results with those we got here. The similar studies could also be done for other non-conventional geometries such as warped $AdS_3$ black holes \cite{Ghodrati:2019bzz}, or in various other massive theories such as New or topologically massive gravity theories \cite{Ghodrati:2016ggy}.

\section{Conclusion}
\label{s6}

The aim of this work was to study the novel methods of detecting the event horizons of black holes specially for the case of massive solutions of massive gravity theories which could specifically be used in numerical relativity solutions. Note that, the non-local notion of the event horizon, which requires information on the future null infinity, seems to be inconsistent with the very first postulate of locality from which general relativity has been constructed. As a matter of fact, this procedure would require to know the entire spacetime evolution for being able to claim whether it contains or not a black hole.

A first proposal for tackling this issue was actually the introduction of the concept of {\it apparent horizon} which is quasi-local, but it is both foliation-dependent and observer-dependent \cite{Ashtekar:2004cn}, and then these ambiguities would affect numerical simulations in which a sharp location of the horizon would be required \cite{Booth:2005qc}. Therefore, it has been conjectured that the horizon should be defined locally as the locus of the zeros of some appropriate ``curvature quantity'' \cite{Coley:2017woz}. Along this line of thinking, one specific proposal relies on the so-called Cartan curvature invariants, which are the components of the curvature tensor and of its derivatives in the canonical frame, as they are also foliation-independent.

One should note that the precise information about the location of a black hole horizon are not only important in light of the previously mentioned simulations of binary black hole systems for extracting the gravitational wave spectra, but also because of  radiation and particles' emission phenomena \cite{Pereira:2015pva}.


In fact, the particle emission process affects the amount of energy which could be extracted from a black hole, and which has been quantified through the Christodoulou-Ruffini irreducible mass formula. The energy emitted by the black hole could then be detected as gravitational radiation \cite{Hawking:1971tu,Blanchet:2006zz,Buonanno:2006ui,Shibata:2011jka,McWilliams:2010eq}, and/or gamma ray bursts \cite{Punsly:2001ng,Ruffini:1999kp}, especially from the active galactic nuclei \cite{Rees:1984si,Begelman:1984mw,King:2008au}.

Since the Christodoulou-Ruffini irreducible mass is sensitive to the physical parameters describing a black hole, like its mass, angular momentum, and electric charge, it has been argued that it may be a valid tool for testing the no-hair conjecture, and the existence of possible extra degrees of freedom beyond the standard model of elementary particles like axions \cite{Arvanitaki:2009fg,Arvanitaki:2010sy}, or bosons as in superradiance phenomena \cite{Press:1972zz,York:1983zb,Sanchis-Gual:2015lje,Rosa:2009ei}, or again in the spectra of gravitational waves \cite{Baumann:2018vus,Brito:2017zvb}. 

Furthermore, the Christodoulou-Ruffini irreducible mass may provide as well new insights into the thermodynamics of black holes because it evolves in time like the horizon area, or the entropy \cite{Bekenstein:1972tm,Hawking:1982dh}.

While the correspondence between surface effects in black hole physics and in ordinary objects has been widely investigated, the idea of irreducible mass and its connection to the processes of particle absorption seems to have been exploited very little, specifically beyond the aforementioned astrophysical applications \cite{Gwak:2015sua,Zhang:2016tpm,Hu:2019zxr}. However, laboratory experiments dealing with analogue black holes may constitute useful environments for detecting a sort of Hawking radiation \cite{Barcelo:2018ynq}, and the loss of information in black hole evaporation \cite{Chen:2015bcg}, while there are no available techniques for observing these astrophysical phenomena in real world.

Therefore, it may turn out possible to investigate the modifications from general relativity, the role of a massive graviton clarified in this paper being an example, and the no-hair theorem in such laboratory settings considering the Christodoulou-Ruffini estimations on the amount of energy which can be extracted from a black hole. In fact, experimentally the black hole horizon can be identified through tidal effects which can be measured using gravimeters.

Keeping all these aspects in mind, in this paper we have actually confirmed the validity of the geometric horizon conjecture in the ``massive gravity BTZ black hole'', and we have provided some curvature constraints which invariantly could quantify the role of the mass of the graviton field. 

Then, we have introduced the  \textit{ \lq\lq massiveness''} parameter, which could act like a ``measure'' of distance between the general relativity and massive gravity. We studied this quantity in different setups and provided various plots. Then, we have investigated the black hole parameters affecting the size of the area of the black hole horizon. Interestingly, we have noted that the term proportional to the mass of the graviton fields could effectively be re-absorbed into the stress-energy tensor entering the field equations, while this is not the case for the cosmological constant parameter, contrary to the case of cosmology in which the latter is interpreted as a dark energy fluid. 

In fact, in lower-dimensional gravity certain curvature invariants could detect the horizon of a black hole only for non-vacuum spacetimes, and we have shown that an electric Maxwell field and the gravitons can provide such matter content, while this is not the case for the cosmological constant.

 \section*{Acknowledgement}

The authors, M.G. and D.G. acknowledge supports from China Postdoctoral Science Foundation with grant numbers 2019M661945 and 2019M661944 respectively.

 {}

\end{document}